\documentclass[compsoc,conference,a4paper,10pt,times]{IEEEtran}

\IEEEoverridecommandlockouts
\usepackage{amsmath,amssymb,amsfonts}
\usepackage{algorithmic}
\usepackage{graphicx}
\usepackage{textcomp}
\usepackage{bmpsize}
\usepackage{xcolor}
\usepackage{lipsum}
\usepackage[colorlinks=true,urlcolor=black]{hyperref}
\def\BibTeX{{\rm B\kern-.05em{\sc i\kern-.025em b}\kern-.08em
    T\kern-.1667em\lower.7ex\hbox{E}\kern-.125emX}}
\usepackage{bbding}
\usepackage{natbib}
\usepackage{pifont}
\usepackage{amssymb}
\usepackage{enumitem}
\usepackage{titlesec}
\usepackage{xtab}
\usepackage{ulem}
\usepackage{balance}

\usepackage{booktabs}

\usepackage{diagbox} 
\usepackage{colortbl}
\usepackage{graphicx}
\usepackage{makecell}
\usepackage{longtable}
\usepackage{multirow}
\usepackage{fourier}

\usepackage{fontawesome}
\usepackage[absolute]{textpos}

\usepackage{pgf}
\usepackage{xcolor}

\usepackage{tikz}
\usepackage{url}
\usepackage{hyperref}
\usepackage{amsmath}
\usepackage[utf8]{inputenc}
\usepackage{multirow}
\usepackage{lscape}
\usepackage{rotating} % rotate table 90
\usepackage{tabularx}
\usepackage{cleveref}
\newcolumntype{C}[1]{>{\centering\let\newline\\\arraybackslash\hspace{0pt}}m{#1}}
\newcolumntype{P}[1]{>{\centering\arraybackslash}m{#1}}

\def\BibTeX{{\rm B\kern-.05em{\sc i\kern-.025em b}\kern-.08em
    T\kern-.1667em\lower.7ex\hbox{E}\kern-.125emX}}

\newcommand{\comment}[1]{}

\usetikzlibrary{arrows.meta}
 
\newcommand*\halfcirc[1][0.75ex]{%
  \begin{tikzpicture}
  \draw[fill] (0,0)-- (90:#1) arc (90:270:#1) -- cycle ;
  \draw (0,0) circle (#1);
  \end{tikzpicture}}
\newcommand*\fullcirc[1][0.75ex]{\tikz\fill (0,0) circle (#1);}
\usepackage{xspace}
\newcommand{\eg}{e.g.,\xspace}
\newcommand{\ie}{i.e.,\xspace}
\newcommand{\ea}{{et al.}\xspace}
\newcommand{\swsc}{software supply chain\xspace}

\newcommand{\model}{AStRA\xspace}

\newcommand{\Paragraph}[1]{\noindent\textbf{#1}}

\newif\ifcommentary
\commentarytrue
\ifcommentary
\newcommand{\todo}[1]{{\color{red}\textbf{TODO:} #1}\xspace}
\newcommand{\fixme}[1]{{\color{red}\textbf{!! FIXME: #1 !!}}}
\newcommand{\msm}[1]{[{\color{orange}msm: #1}]}
\newcommand{\Eman}[1]{[{\color{blue}Eman: #1}]}
\newcommand{\santiago}[1]{[{\color{green}Santiago: #1}]}
\newcommand{\Juanita}[1]{[{\color{brown}Juanita: #1}]}
\newcommand{\Alvaro}[1]{[{\color{magenta}Alvaro: #1}]}

\else
\newcommand{\todo}[1]{}
\newcommand{\fixme}[1]{}
\newcommand{\msm}[1]{}
\newcommand{\Eman}[1]{}
\newcommand{\santiago}[1]{}
\newcommand{\Juanita}[1]{}
\newcommand{\Alvaro}[1]{}
\fi

\newcommand{\greybox}[2]{
\noindent
	\begin{minipage}{\columnwidth}
    \vspace{3pt}
   \fcolorbox{black}{gray!10}{%
    \begin{minipage}{\columnwidth-1em}
  \textbf{#1:} #2
   \end{minipage}%
  }
  \vspace{3pt}
\end{minipage}
}

\newcounter{insights}
\newcommand{\insightbox}[1]{
	\greybox{Insight~\number\value{insights}}{#1}%
	\stepcounter{insights}
}

\begin{document}

\title{SoK: A Defense-Oriented Evaluation of Software Supply Chain Security }

\author{\IEEEauthorblockN{Eman Abu Ishgair }
\IEEEauthorblockA{\textit{Electrical and Computer Engineering} \\
\textit{Purdue University}\\
West lafayette, USA  \\
eabuishg@purdue.edu}
\and
\IEEEauthorblockN{Juanita Gomez}
\IEEEauthorblockA{\textit{Computer Science and Engineering} \\
\textit{University of California, Santa Cruz}\\
Santa Cruz, USA \\
jgomez91@ucsc.edu}
\and
\IEEEauthorblockN{Marcela S. Melara}
\IEEEauthorblockA{\textit{Corporate Technology Office} \\
\textit{Intel Corporation}\\
Hillsboro, USA \\
marcela.melara@intel.com}
\and
\IEEEauthorblockN{Alvaro Cardenas}
\IEEEauthorblockA{\textit{Computer Science and Engineering} \\
\textit{University of California, Santa Cruz}\\
Santa Cruz, USA \\
alacarde@ucsc.edu}
\and

\IEEEauthorblockN{Santiago Torres-Arias}
\IEEEauthorblockA{\textit{Electrical and Computer Engineering} \\
\textit{Purdue University}\\
West lafayette, USA  \\
santiagotorres@@purdue.edu}
}

\maketitle
\begin{abstract}
The software supply chain is a highly complex ecosystem of tools, workflows, and human factors involved in creating a piece of software.  Due to its central role in critical systems, it has become a frequent target of sophisticated attacks. Prior work has focused primarily on classifying attack vectors, but there is a gap in discussing and understanding effective defenses. This paper proposes a defense-oriented systematization of software supply chain security that models the fundamental components and causal relationships. Based on 76 works from academia and industry, we define key security objectives and defense patterns of established \swsc defense techniques.
The goal of our taxonomy is to identify major opportunities for future research, and aid \swsc practitioners when
reasoning about the effectiveness of defenses available today.

%in needed to mitigate common attacks and propose a dual taxonomy that systematizes both academic research and real-world tools based on their ability to address these objectives.
%While previous surveys have mostly focused on academic literature surrounding software supply chain security, there has been a notable gap in discussing the practical tools deployed in real-world software supply chain processes. 
%This framework enables a structured evaluation of defense coverage throughout the software supply chain. We validate our approach against known attack taxonomies and highlight research gaps, proposing directions for building secure-by-design software development tools and systems.
%This paper provides a dual taxonomy: one categorizing academic works on software supply chain security, and the other detailing the tools employed within various stages of the software supply chain process. By contextualizing contemporary tools within the software supply chain framework, we offer a comprehensive overview of both theoretical insights and practical applications. While previous surveys have mostly focused on academic literature surrounding software supply chain security, there has been a notable gap in discussing the practical tools deployed in real world software supply chain processes.
\end{abstract}

\begin{IEEEkeywords}
Supply chain security, Computer security, Risk analysis, Software engineering, Data models, Authentication, Integrity, Provenance
\end{IEEEkeywords}

% \begin{itemize}
% \item finish and simplify tables
% \item resolve comments
% \item Major lines:
%     - cohesion 
%     - simplification
%     - compelling
% \item generate list of things we can conclude to put in final section 10
% \item Pass over section 4 and write its discussion (Santiago)
% \item Should we include OSS vs Corporate CS environments
% \item Is there anecdotal evidence, but not a systematic study of these costs?

% \end{itemize}

\section{Introduction}

%Modern software is only as secure as the extensive supply chains that create, compile, package, and distribute it. 

%Because even well-resourced organizations depend on thousands of third-party components, protecting the end product demands defenses that extend far beyond traditional code-centric vulnerability scanning. 

Recent high-profile
attacks on the software supply chain (\eg~\cite{ilgayev_how_2022, logfourj, pypi-typosquatting, sgx-signing-injection, solarwinds-fireeye, 
php-source-code,xz-redhat}) have motivated governments~\cite{cyber-eo, eo-memo, eucyber} and industry associations to develop frameworks to audit the Software Development Life Cycle (SDLC) and validate open source and proprietary software (\eg~\cite{in-toto, iso-sec-dev, nist-app-vetting, oscal, scitt, scorecards, sec-sw-factory, sigstore, slsa, spdx}).

Although the research community has developed detailed taxonomies of open-source software supply chain \textit{attacks} (e.g.~\cite{ohm-backstabbers, Github-CI-Security, krohmer-scored, ladisa-java, ladisa-sok, maloss, npm-analysis, sec-ci, cappos-mirror-2008}), this adversary-centric perspective leaves practitioners with two key challenges. First, these taxonomies rarely specify the \textit{security objectives} defenders should pursue to counter identified threats. Second, they focus on isolated stages of the supply chain, such as package repositories or CI pipelines, ignoring the \textit{transitive and interconnected nature of modern ecosystems}, where a single weakness can cascade throughout the chain. As a result, the defense landscape is fragmented: administrators scramble to adopt point solutions (SBOM generators, secret scanners, sandboxed builders, etc.) without clear guidance on how these tools interoperate or where critical gaps remain.

%Although the research community has produced detailed taxonomies of open source software supply chain \textit{attacks} (\eg~\cite{ohm-backstabbers, Github-CI-Security, krohmer-scored, ladisa-java, ladisa-sok, maloss, npm-analysis, sec-ci, cappos-mirror-2008}), this adversary-focused lens leaves practitioners with two unresolved challenges.  First, attack taxonomies do not identify the \textit{security objectives} that a defender should strive to neutralize those threats.%; engineers are left to infer what to build or buy.  
%Second, prior studies often examine isolated stages (\eg package repositories or CI pipelines) in a vacuum, overlooking the complex, transitive relationships that let a single weakness propagate across an entire chain.  The result is a fragmented landscape in which vendors race to adopt point solutions (SBOM generators, secret scanners, sandboxed builders, \textit{etc.}) without clear guidance on how these pieces fit together, or where critical gaps remain.

%\Paragraph{Positioning Relative to Prior Taxonomies.} 

In this paper, we focus on a defense-oriented systematization of \swsc security. We introduce \textbf{AStRA}, a graph-based model that treats a supply chain as a directed acyclic graph of \textbf{princip\underline{A}ls}, \textbf{\underline{St}eps}, \textbf{\underline{R}esources} and \textbf{\underline{A}rtifacts}. We then perform a systematization of \swsc defenses from academic papers and the gray literature, classifying each defense by \textit{what} it protects, \textit{how} it works, and \textit{where} it fits in the supply chain DAG.  Our bottom-up analysis reveals hidden interdependencies between defenses and allows us to derive a set of fundamental \textit{security objectives} whose aim is to contain most documented \swsc attacks. Further, we highlight objectives with limited coverage and map them to cross-cutting \textit{defense patterns} that describe how a defense technique meets its security objective.

As such, our paper builds on methods in~\cite{Okafor2022} and
complements attack-oriented work~\cite{ladisa-sok,ohm-backstabbers}
by providing an objective-driven framework to understand and apply targeted defenses throughout the software supply chain.

\Paragraph{Positioning Relative to Prior Taxonomies.}
While attack-oriented taxonomies provide valuable insight into how systems are compromised by classifying adversary techniques and vectors, they do not explicitly define the security properties that defenses must guarantee, leaving practitioners to infer what protections are required.

In contrast, our work takes a defense-oriented approach by deriving a set of security objectives that specify the guarantees a secure supply chain must satisfy, independent of specific attacks. This enables reasoning not only about individual mechanisms, but also about coverage and gaps across the supply chain as a whole. 

Prior work has explored related ideas in limited contexts. for example, defining workflow-specific objectives~\cite{gha-analysis} or enumerating safeguards~\cite{ladisa-sok}, but these efforts either focus on particular environments (e.g., CI systems) or catalog mechanisms rather than formalizing cross-cutting objectives. 
As a result, guidance in practice is often expressed as collections of best practices (e.g., signing, MFA, sandboxing), without clearly specifying the security guarantees they are intended to provide. We instead distinguish between mechanisms and objectives: best practices (e.g., signing, MFA, sandboxing) are mechanisms that describes how to defend, independent of implementation, whereas security objectives define properties that must hold (e.g., non-equivocation). Best practices are often technology- and implementation-specific and may evolve as new tools emerge. In contrast, security objectives capture guarantees that can be realized through different defense mechanisms over time. This abstraction also allows for flexibility when choosing techniques based on operational context if they satisfy the same underlying guarantees.

%, we hope to help researchers, practitioners, and policy makers prioritize investments and requirements to improve the security of their supply chains. 
%Our ultimate goal is to establish a preliminary objective-driven framework that
%can aid security researchers and software vendors alike in reasoning about their
%\swsc security posture when adopting or developing new defense techniques.
%\Alvaro{Delete next sentence? It is not connected to previous sentence, and seems redundant as these papers were mentioned before}As such, we include a comparative analysis between our security objectives and the Ladisa~\ea SoK~\cite{ladisa-sok}---the current state-of-the-art attack taxonomy---establishing an explicit mapping between attack classes and desired security objectives to mitigate each class of attacks.

% Further, our work examines techniques from other areas of computer science and
% information security that enable resilient-by-design technological and human-centered defenses, and recommend new research opportunities.
%% such as systems, PL/formal verification, usability, CS education, hardware design, which is done in software.
%Ultimately, we aim to provide a guide for the academic community to help advance the state-of-the-art defenses.

%% \msm{where does this fit?}
%% no good understanding of what works generically vs domain-specific (e.g. IoT vs cloud supply chain)

\noindent Our main contributions include:
%\begin{enumerate}
%    \item A taxonomy of academic research on supply chain security. 
%    \item A categorization of security tools used within supply chain processes.\Eman{Are we keeping 2?}
    %\item a graph-based conceptual framework for holistically representing a software supply chain
%    \item identification of security objectives for each \swsc element, and a taxonomy of defense approaches (\S\ref{sec:principals}-\S\ref{sec:topology});
    %\item {an in-depth validation of the AStRA model}
 %   \item a discussion of research gaps and future directions (\S\ref{sec:conclusion}).
%\end{enumerate}

\begin{itemize}
  \item \textbf{AStRA, a principled model of the software supply chain.} We formalize four core elements and their causal relationships, allowing more precise representation of and reasoning about transitive risk.
  \item \textbf{An objective-driven defense taxonomy.} We derive fifteen foundational security objectives through systematic evaluation of prevalent academic and industrial mechanisms, and their properties.
  \item \textbf{A unified systematization of defenses.} Our analysis spans 76 techniques (e.g., transparency logs, threshold signing) and specialized tools (e.g., container isolation, provenance attestations), revealing both overlaps and blind spots.
  \item \textbf{Actionable insights and research directions.} We identify underexplored areas, such as DAG-level integrity, usable signing workflows, artifact confidentiality, and AI-generated code provenance, and outline concrete opportunities for the community to close these gaps.
 \item \textbf{New scripts and datasets.} We also make publicly available the scripts we used to collect the papers used in our systematization and our collected results \url{https://github.com/TSELab/astra-artifacts} %We hope that these resources help inspire new research in \swsc and expand our results. 
 \end{itemize}

\section{Systematization Methodology}
\label{secsec:sok-structure}
%This paper discusses and classifies the papers related to supply chain security.
%We start by collecting and categorizing work in the software supply chain security space. 
%We do this with the goal of identifying trends, gaps and sub-areas of research in the space.

Our methodology consists of two primary phases:
1) automated collection of academic databases using keywords and a list of venues, and
2) a multiple-step review and labeling process by at least two coauthors of this paper.

% STA: I changed "scraping" to "collection" because I want to focus on the outcome
% *NOT* on the mechanism. E.g., if there was an API for this, then we wouldn't 
% need to scrape
\subsection{Phase 1: Paper collection}

% STA: Here, I tried to focus on convicing the reader we were following a 
% structured approach that's "blessed" by the community
%We collected papers through various stages.
Following guidelines for systematic literature studies~\cite{wohlin_guidelines_2014}, we collect a seed of papers using keyword search and then use this seed to further expand the corpus of work to study.
To collect as much relevant work as possible in this first phase, we filter the work in the second phase, so we look for the term "supply chain" and variations of it like "supply-chain" in security conferences. 

To ensure high quality and impactful research, and given that an open query would result in an unmanageable number of papers (More than 1,850,000 results as of 2025), we constrain our search to publications within established computer security venues. We seed our search with system security conferences with large citation rates (citations/accepted papers), participant rates (registered participants/accepted papers), and low acceptance rates, as summarized in \cite{zhou_conference_ranking}. We list the selected conferences in Table~\ref{tab:conferences} in App.~\ref{app:methodology}. We then use Google Scholar's API and search for the conference sources using keywords such as \texttt{"supply chain" source:"USENIX Security Symposium"} as a query to filter supply chain work presented at the USENIX Security Symposium. For each paper, we extract the title, authors, year, venue, abstract, citations, and URL. The results of this search, conducted between February and June 2025, are compiled into a single dataset of 671 potential software supply chain security papers. The collected papers can be found in our shared repository \url{https://github.com/anonymous-eurosp/astra-artifacts}.

An interesting anecdote from these results is that we found citations that appear to be hallucinated by an AI agent.
%, which presumably helped some authors write recent papers (i.e., recent papers cite a non-existent paper and then Google Scholar picks %up this hallucinated citation and adds it to their entries).
In particular, Google Scholar has entries with ``supply chain" in the title but has no direct link to the proceedings of the conference, and when we looked at the proceedings of that year, we did not find the collected publication. 

% STA: minor edit to instead refer to other sources to indicate how we 
% choose venues
%the top-tier and selected tier-2 security conference venues. These venues include: 

\subsection{Phase 2: Manual Filtering and Scope}

Phase 1 resulted in more than 100 papers on the AI supply chain and the hardware supply chain. We deliberately scope this study to the software supply chain to provide a focused systematization of the defenses within this rapidly evolving and practically impactful domain. Although related areas such as the AI supply chain and the hardware supply chain raise equally important security concerns, including data poisoning, model tampering, or hardware trojans, each involves distinct threat models, stakeholders, lifecycle processes, and technical challenges. These domains merit their own in-depth systematization efforts to account for their unique complexities and to avoid conflating fundamentally different ecosystems under a single framework. Our goal is to provide clarity and structure to the software supply chain security landscape while encouraging future but complementary SoK work that can do the same for AI and hardware supply chain defenses. 

Therefore, we perform various filtering passes, ensuring that each paper is reviewed by at least two authors who agree on the following questions:
% STA: below, I tried to give examples to provide a more concrete sense of how we are applying our criteria
\begin{enumerate}
    \item Does the paper focus on the software supply chain, rather than physical, hardware, or AI supply chains?
    \item Is software supply chain security a central topic of the paper, rather than a peripheral motivation (e.g., a security challenge) or a side note?
%focusing on Machine Learning or Artificial Intelligence supply chain were marked as ``not relevant''
%Papers primarily focused on Machine Learning (ML) or Artificial Intelligence (AI) security in a general context, without a specific connection to the software supply chain, were marked as "not relevant."
%    For example, various papers on trojanizing machine learning models utilize a software supply chain attack as a possible avenue for an attacker to introduce a ML trojan.
\end{enumerate}

% I wanted to focus this portion on the goal of the paper, so instead of 
% talking about tangential benefits (categorizing the work of other types of 
% sc), we categorize all and we provide general statistics.
 Disagreements between the reviewers were resolved through discussion, with a third coauthor serving as tie-breaker if needed to ensure consistency and minimize bias.
Of the 671 papers from Phase 1, 98 of them are marked as ``relevant'' by both reviewers. We then classify these papers according to their core contribution to supply chain security. 
We identify four main categories as described below (see Table~\ref{tab:cat-refs}). 
\begin{itemize}
        \item \textbf{Attack:} If the primary contribution of the paper is attacks on software supply chains.%~\cite{impersonation,roberts2023adversarial,file-hijacking,haste-speed,ghunter,trojan-source,union-duress,microcoded-processors,ohm-backstabbers}. \Alvaro{testing to see how the table would look here instead.}
        \item \textbf{Defense:} If the main contribution of the paper is a new or newly applied defense mechanism.
        \item \textbf{Analysis:} If the paper evaluates the effectiveness of \textit{existing} security controls, policies, or practices within the supply chain by conducting a measurement study (e.g., the extent to which Android applications update their third-party libraries) or analyzing the security of existing defenses (e.g., using formal methods to evaluate a defense).
        \item \textbf{Survey:} If the paper provides a review, taxonomy, or Systematization of Knowledge for existing work in the area of software supply chain security.%\cite{sok-docker,ladisa-sok}.
\end{itemize}

\begin{table}[ht]
  \scriptsize
  \setlength{\tabcolsep}{3pt}      % tighter column spacing
  \renewcommand{\arraystretch}{1.15}
  \caption{Categorized academic references (filtered + extended).}
  \label{tab:cat-refs}
  \begin{tabularx}{\columnwidth}{@{}lXc@{}}
    \toprule
    \textbf{Category} & \textbf{References} & \textbf{Count} \\ \midrule
    Defense &
      \cite{karthik_uptane_nodate, bytes_fieldfuzz_2023, noauthor_natisand_nodate, Moore2023, ladisa_feasibility_2023, nguyen_up2dep_2020, ntousakis_demo_2021, noauthor_hodor_nodate, Newman2022, merrill_speranza_2023, vu_towards_2020, noauthor_poster_nodate, gobbi_poster_2023, HARP, dunlap_vfcfinder_2024, huang_vmud_2024, wyss_wolf_2022, wen_5g-spector_2024, meng_identifying_2017, noauthor_product_nodate, noauthor_argus_nodate, chainiac, christensen_decaf_2020, huang_donapi_nodate, pantelaios_fv8_nodate, he_rapidpatch_nodate, han_sigl_nodate, woo_v1scan_2023, TorresArias2019, wu_your_nodate, zhou_hey_2025, Anomalicious, verifiability, Sofaer_RogueOne, Jiang_BinAI, Boughton2024, Sun_where, Luo2024, Jia2024, jarhead, spider, contour_2018, sec-ci, guix, dettrace, Policy-Transparency} & 46 \\ \midrule
    Attack &
      \cite{impersonation,roberts2023adversarial,file-hijacking,haste-speed,%
      ghunter,trojan-source,union-duress,microcoded-processors,ohm-backstabbers} & 9 \\ \midrule
    Security Analysis &
      \cite{uptane-analysis, manifest, rapid-api, kloeg2024charting, android, jack-in-the-box, keep-me-updated, plugins-security, Wermke2023, reproducile_builds_usability, android-analysis, committed-to-trust, intrusion, everyone-for-themselves, package-security, package-signing, trust-verify, crypto-detectors, sandriller, typosquatting, bilingual, Github-CI-Security, mistrust, pushed-by-accident, unveiling-iot, unveiling, seum-spread, manufacturing-industry, Zahan2023a, secure-sgx, npm-analysis, reproducibility, Stalnaker2024, Malka2024, Pan_ppt4j, Herzig2022, Pinckney2023, Zahan2024, Jungwirth2023, Ponta2019, Delicheh2024} & 41 \\ \midrule
    Survey & \cite{sok-docker,ladisa-sok} & 2 \\ \bottomrule
  \end{tabularx}
\end{table}

\Paragraph{Extending paper selection.}
We select the 46 defense papers identified in phase 2, and enrich our paper selection by consulting industry forums, guidelines, open source recommendations, and government requirements. 
However, a systematic aggregator of gray literature is not available, so we rely on the author's engagement with industry and government to select adequate texts.
Specifically, we selected work from the Internet Engineering Task Force (IETF)~\cite{scitt}, the U.S. National Institute of Standards and Technology (NIST)~\cite{nist-800-204D}, the Center for Internet Security (CIS)~\cite{cis-sscs-guide}, the Linux Foundation~\cite{cncf-best-practices}, and large open source communities~\cite{srlc-linux-handbook, PEP480}.
We also extend our selection of papers to include proposed systems (these may also be published in top systems and networking conferences) if they are mentioned in these guidelines.
Similarly, we identify tools and frameworks that instantiate properties outlined in these guidelines. 
This adds 6 papers from industry literature, and 24 tools that were not introduced via academic work, yielding a total of 76 defense-oriented works. 
 The table containing these additional works can be found in Table~\ref{tab:appendix_extra_works} in the appendix.

%\msm{We should list these systems and industry in a table in the appendix. BTW in table 4 (our master table) I only counted 23 tools.}

As the field of \swsc research has matured, we can see that surveys and SoKs started in 2023. As mentioned above, a previous SoK focused on attacks~\cite{ladisa-sok}, and a more recent one~\cite{sok-docker} focuses only on docker attacks and defenses. Our goal is to complement these previous efforts by focusing on defenses against a wide range of supply chain threats. We elaborate on trends in the distribution of \swsc research references in App.~\ref{app:methodology}.

\section{Modeling the Software Supply Chain}
\label{sec:model}

%\Alvaro{tentative references for dependency graphs \cite{small_world,npm-analysis,maloss,keep-me-updated,typosquatting,cappos-mirror-2008,union-duress} and for flow graphs \cite{noauthor_argus_nodate,Scudo,intrusion,in-toto-layout-tool,git-attacks} }

The literature surveyed tells us that modern software supply chains span numerous actors, tools, services, and artifacts distributed across organizational and administrative boundaries. 
Understanding the security of how software is created, transformed, validated, and released requires more than simply identifying individual vulnerabilities; it requires
%a clear representation of the structure and behavior of the entire chain. 
%A precise supply chain model is the foundation for 
evaluating defenses, reasoning about transitive risk, diagnosing compromises, and understanding how different components interact. 
%Without such a model, it becomes difficult to 

Specifically, stakeholders ask \swsc security questions such as: Which elements influence a given artifact? Where are trust boundaries crossed? How does a compromise in one part of the ecosystem propagate to others? And how do different defenses address key security goals?

Addressing these questions requires a precise model of the structure and behavior of the supply chain. Graph-based representations help capture the relationships among code dependencies, build steps, and human actors, enabling the analysis of transitive risks and potential attack paths. Two such representations dominate the literature: dependency graphs and flow graphs.

\textit{Dependency graphs}
%represent artifact-adjacency properties. 
map the relationships between software components based on reliance and consumption at a single point in time $t$. 
In a dependency graph, the nodes represent packages or components, and the edges represent ``artifact adjacency'', that is, one artifact depends on or includes another~\cite{small_world,npm-analysis,maloss,keep-me-updated,typosquatting,cappos-mirror-2008,union-duress}. 
Of note, dependency graphs are \textit{speculative}, for most dependencies are not \textit{resolved} until an artifact propagates through the chain.
A good example is package dependency networks, which rely on a package manager to carry out \textit{dependency resolution} to identify the best candidates at the current time $t$. 

In contrast, \textit{flow graphs} focus on representing the sequence of transformation steps that result in the creation of a software artifact -- often likened to a CI/CD pipeline or development lifecycle~\cite{noauthor_argus_nodate,Scudo,intrusion,in-toto-layout-tool,git-attacks}. 
Here, nodes represent build steps, processes, or stages (such as compile, test, package, deploy), and directed edges represent the sequence or data flow between these stages.

%However, we find these abstract models insufficient as they do not characterize the relationships between artifacts and steps (e.g., is an artifact used as a tool or as an input to a tool?) and often leave out other core supply chain elements. Dependency graphs simplify relationships and actors, leading to blind spots and forcing security efforts to remain local and reactive. Similarly, flow graphs cannot capture the relationships between artifacts and steps; for example, not distinguish if an artifact is merely consumed as an input or if it is used as a tool (like a build compiler or a training framework), which affects the integrity of the output. An attack may exploit elements omitted from these simple graph types. 
%\Alvaro{this paragraph needs to be stronger: show a specific example of why these two previous models fail... e.g., in modeling an attack; and then suggest that something is missing: Causality. Define it properly and then show why previous models can't capture it.}

However, these abstract models are insufficient for reasoning about supply chain security because they fail to capture how actions taken by one element \textit{influence} the next. 
Consider an attack on a CI/CD pipeline where a compromised build runner injects malicious code during compilation. 
A dependency graph cannot represent this scenario: artifact adjacency remains unchanged -- the same package depends on the same libraries -- yet the resulting artifact is malicious. 
A flow graph performs only slightly better: it captures the ``build step'' but cannot express which resource carried out the step, which principal authorized it, or why the resulting artifact deviated from expectations. 

Both models lack the notion of \textit{causality}: i.e., they cannot represent that a specific action, performed by a specific actor using a specific resource, directly or \textit{indirectly}, caused a compromise.
%the introduction of a malicious artifact. 
Without reasoning about causality, neither model can explain how a compromise propagates: whether an attack originated from a poisoned dependency, a misconfigured CI service, a stolen credential, or a removed validation step. 
As shown in other domains, such as root-cause analysis for intrusion detection~\cite{batesRCAID}, this missing causal structure is precisely what prevents these models from supporting post-mortem analysis, transitive risk reasoning, and principled defense evaluation.

\subsection{A Causal Supply Chain Model}
\label{secsec:astra-model}

To address these limitations, we introduce a new causality model for the supply chain, which we call AStRA. 
AStRA treats the entire software supply chain as a directed acyclic graph (DAG) of four element types: \textbf{Principals} (e.g., people or organizations), \textbf{Artifacts} (e.g., code or artifacts produced/consumed), \textbf{Resources} (e.g., tools, services, infrastructure), and \textbf{Steps} (e.g., actions or operations).
Edges in this graph represent causal links: Principals use Resources; Resources carry out Steps; and Steps consume input Artifacts and produce output Artifacts. 
This forms an actor-artifact graph that richly encodes who did an action with what resource, leading to what artifact. 

This causal graph expands traditional dependency graphs, which only show artifact-to-artifact relationships, and beyond flow graphs, which show step sequences without tying in the human or resource responsible for an action. 
By including actors and distinguishing ``active'' nodes (such as build systems, signing services, compilers) from ``passive'' nodes (code, binaries, configurations), AStRA can represent the context relevant to security that earlier models miss. 
For instance, in-toto’s~\cite{TorresArias2019} layout model records steps and their materials, but does not explicitly differentiate whether a given input was a piece of source code or a compiler program – AStRA makes that distinction clear by labeling the latter as a Resource.

%clarity is essential for causality: 
When evaluating the design and effectiveness of \swsc defenses, understanding how a malicious artifact, for instance, can be introduced is essential (via a compromised dependency artifact or via a compromised build tool resource?). 
%The AStRA graph structure enables us to trace these cause-and-effect chains throughout the supply chain DAG.
Thus, we leverage AStRA to systematize \swsc defenses in a way that makes causality explicit. This model will also help us systematize the defenses. 

\begin{figure*}
    \centering
    \includegraphics[width=.9\textwidth]{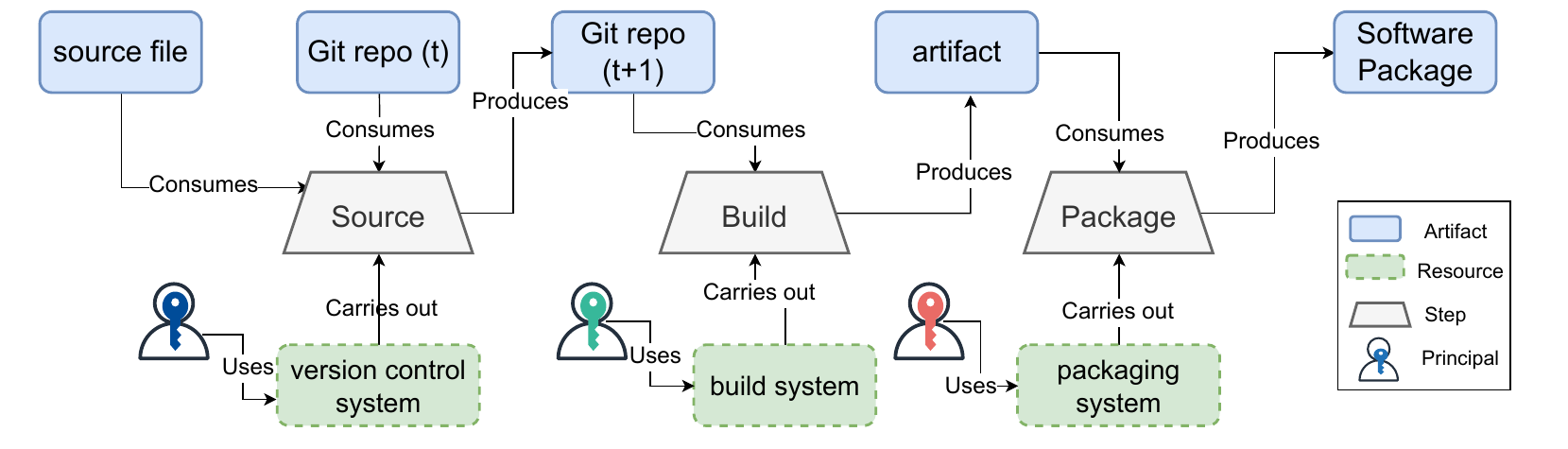}
    \caption{Software supply chain DAG structure example that consist of three standard steps: sourcing, building, and packaging.}
    \label{fig:topology}
\end{figure*}
\Paragraph{AStRA example.}
Figure \ref{fig:topology} shows a \swsc DAG structure that covers three common steps:

\textbf{Source} is the initial phase, which involves the creation and editing of the source code for the final artifact. This code is typically stored and managed in a repository hosted on platforms such as GitHub~\cite{github_2025}.

\textbf{Build} refers to the process of transforming input artifacts (source code) into output artifacts (e.g., a binary). The resources used to carry out this step often consist of a
developer's local machine or a cloud-based CI/CD service.

\textbf{Package} is often the final step in which the output of a build process is ``packaged'' to facilitate ``publishing'' the final artifact for use by others. 
Package artifacts can be found in ecosystem-specific package managers like PyPI (Python Package Index) or NPM (Node Package Manager). 

Note that more complicated structures often exist and may include steps to validate the behavior of an artifact (\eg by means of running a test suite on a binary), or perform
security-related tasks (\eg code signing).
However, all steps carried out in the chain can be represented as a specific DAG structure of the elements introduced above.

To illustrate the practical application of our framework, we use the scenario in  Figure \ref{fig:topology}  as a running example throughout this paper. This scenario represents a Containerized Microservice Build Pipeline. In this context, the Principal is represented by a Developer (Alice), the Steps include 'Compile' and 'Containerize', the Resources consist of the GitHub Actions Runner, and the resulting Artifacts are the Source Code repository and final Docker Image.

We validate AStRA's fidelity and generality in  Appendix~\ref{sec:evaluation} (as well as showing how AStRA can be used to model popular supply chain failures). In particular we show an application of our tool to model the supply chain for 990 debian builds in Figure~\ref{fig:debian-astra}.

%Lastly, while it may be tempting to use edge properties to capture relationships (e.g., whether a node is an artifact or a resource in a step), this will only introduce complexity for the model without an immediate benefit. \msm{this isn't super clear to me. I'd argue (and have gotten feedback from researchers familiar with astra) that it's already not a trivial model.}

%Understanding these st is crucial for identifying potential vulnerabilities and securing the software supply chain. Each stage, from source to package, presents unique challenges and opportunities for attacks. %This leads us to the next section, where we describe the specific threat model for software supply chain security.

%Identifying the fundamental elements of a software supply chain enables principled reasoning about the security of individual elements. Yet, evaluating each element in isolation is not sufficient to understand how the causal relationships between these elements impact the overall security of a software supply chain. So, we include a representation of the supply chain structure as a directed acyclic graph (DAG) in which artifacts, steps, resources and principals are vertices and the edges between them represent their relationships. 

%\Alvaro{No threat model?}
\subsection{Threat Models}
% \Paragraph{A few special cases.}
% For the sake of distinguishing between artifacts operated on during a step from tools (\eg a compiler) used to \textit{perform} a step, we consider tools to be resources
%Our AStRA model can help us systematize defenses to the \swsc by identifying 
%for every element , define how it can be compromised, and why/what can he do?

%\msm{a lot of the citations in here seem ``random'' and don't appear in table 1, we should probably do a pass to replace some of these with papers that we collected.}
By formalizing these four elements, we can precisely define the threat landscape. We summarize the threat model for each element, plus the DAG structure itself:

%\noindent{\textbf{Principals.}}
Threats to \textit{principals} aim to subvert the authentication and authorization process by stealing or reusing credentials %noauthor_certcc_nodate 
~\cite{ohm-backstabbers}, or by exploiting overly privileged roles~\cite{Github-CI-Security}. 
A compromised principal can authorize malicious steps downstream: such as approval of a tampered build~\cite{solarwinds-report}, altering a critical artifact~\cite{php-source-code}, or modifying CI logic while remaining indistinguishable from a legitimate principal~\cite{typosquatting}.%\santiago{cite typosquat decarli here}

%\noindent{\textbf{Artifacts.}}
\textit{Artifacts}  are targets for tampering during creation~\cite{trojan-source}, in transit~\cite{cappos-mirror-2008}, or maliciously substituted after caching or redistribution~\cite{goodin2025gomirror}. 
These attacks exploit the fact that artifacts are frequently reused across packages and systems.

%\noindent{\textbf{Resources.}} 
Threats to \textit{resources} seek to change a resource's intended behavior during a step, or abuse over-privileged execution environments to perform unauthorized actions~\cite{haste-speed}. %,uberti_log4j_2021,noauthor_phpinternals_nodate,noauthor_gentoo_nodatee
Because resources are often reused across different steps and supply chains, compromising them can have wide reaching effects across the supply chain graph~\cite{uberti_log4j_2021}.

%\noindent{\textbf{Steps.}}
Threats to \textit{steps} aim to modify the behavior of a step to operate in an unexpected or unauthorized way like running malicious builds~\cite{solarwinds-report, ohm-backstabbers} or to conceal attacks on related supply chain elements. 
Manipulated step parameters (e.g., altering configuration flags to disable security checks~\cite{ilgayev_how_2022}) 
is a common threat.
%or forging its outputs to misrepresent what was executed. 
Attackers can also compromise a step via any other element, \eg using a compromised compiler resource~\cite{noauthor_apples_2015}. 

%\noindent{\textbf{DAG Structure}}
%The AStRA model allows us to explicitly reason about attacks on the supply chain that exploit their transitive and distributed nature \textbf{via its DAG structure}.
Finally, threats to the \textit{supply chain DAG} aim to
%do not aim to modify existing edges and vertices, but rather 
add or remove vertices or edges to alter the flow and outcome of a software supply chain. 
Real-world examples include removing an artifact from a package manager~\cite{leftpad-npm}, or adding a malicious package during the creation of a Docker image~\cite{birsan_dependency_2023}%\santiago{add dependency confusion here}. 
Another form of structural manipulation omits critical steps. For example, misconfigured GitHub Actions workflows allow untrusted pull requests to trigger a publish step without validation (e.g., tests or reviews)~\cite{Github-CI-Security}, enabling attackers to propagate malicious artifacts.

\insightbox{Each \swsc element (incl. the entire DAG structure) presents an attack vector with a unique threat model. While threats target standard security expectations (\ie confidentiality, integrity, or availability) of the different elements, practitioners and researchers lack a unified framework for reasoning about the effectiveness of defenses against these threats.}

%\msm{I would the evaluation between sections 4 and 5}
%\input{Evaluation}

% \noindent
% \begin{minipage}{\columnwidth}
% \vspace{3pt}
% \fcolorbox{black}{gray!10}{%
%     \begin{minipage}{.97\columnwidth}
%   \textbf{RQ1 Summary:} We can represent the causal nature of the \swsc by \textbf{abstracting} the software supply chain as \textbf{four fundamental elements} and representing them as vertices in a directed acyclic graph \textbf{(DAG)}. % AStRA captures the \textbf{intricate} structure of workflows, encodes their \textbf{causal} relationships, and reveals \textbf{transitive} risk propagation. 
%   This abstraction enables \textbf{systematic reasoning about risk} across heterogeneous supply chain structures.
%    \end{minipage}%
%   }
%   \vspace{3pt}
% \end{minipage}

\section{Defense Systematization}
\label{secsec:systematization}
  
We organize the defenses in our corpus using AStRA (\S\ref{sec:model}), categorizing each technique by the security objective it addresses and the supply-chain element it protects. This structure allows us to compare defenses across objectives and identify recurring patterns that reveal both overlaps and gaps in the current landscape. Table~\ref{tab:sok-table-compact} summarizes the four dimensions of our taxonomy.

\begin{itemize}
    \item \textbf{Defense Technique:} The specific technical mechanism or approach used in the defense; techniques are often well-established in other areas of security, but newer to the supply chain.
    \item \textbf{Security Objective:} The primary security goal that a defense technique is meant to achieve.
    \item \textbf{AStRA Element:} One of five classes of participants in the software supply chain (\S\ref{sec:model}) that a defense targets, \ie principals, artifacts, steps, resources and the overall DAG structure.
    \item \textbf{Defense Pattern:} How the defense implements a particular security objective based on established security control types~\cite{wright_control_types2008}.
% \msm{I think we ought to move these to section \ref{sec:defense-patterns}}
% \begin{itemize}
% \item \textbf{Detective:} Information to help identifying anomalies, unauthorized changes, or misuse through mechanisms like \emph{transparency} (e.g., logging, provenance tracking) that enable continuous auditing.
% \item \textbf{Corrective:} Enabling systems to gracefully handle compromise or failure through \emph{redundancy}, \emph{graceful degradation}, and other \emph{corrective} techniques that support containment, rollback, or continuity.
% \item \textbf{Preventative:} Proactive mechanisms, such as ensuring \emph{correctness}, and applying \emph{least privilege} and other forms of \emph{limitation} to reduce the attack surface.
% \end{itemize}
\end{itemize}

%We present our taxonomy in Table~\ref{tab:sok-table2}. 
% Our systematization discussion begins with defenses for
% principals (\S\ref{sec:principals}). We then discuss defenses for artifacts (\S\ref{sec:artifacts}).%, which represent the unit of information passed between steps.
% Because resources encompass both tool artifacts and services,
% our analysis of defenses for resources (\S\ref{sec:resources}) builds upon the analysis of artifact defenses.
% Similarly, reducing risks to steps (\S\ref{sec:steps}) and the supply chain DAG (\S\ref{sec:topology}) requires 
% a number of unique approaches on top of defenses discussed in prior subsections.

%For each of these elements, as well as the whole DAG structure, we discuss different academic and industry approaches that seek to defend it. For all the discussed approaches, we identify security objectives that the defense resources try to achieve. %(towards \textbf{RQ2}).
To organize our systematization, we enumerate security objectives for each AStRA element: \textbf{P\#} for principals (\S\ref{sec:principals}), \textbf{A\#} for artifacts (\S\ref{sec:artifacts}), \textbf{R\#} for resources (\S\ref{sec:resources}), \textbf{S\#} for steps (\S\ref{sec:steps}), and \textbf{T\#} for the supply chain DAG structure (\S\ref{sec:topology}). Under each security objective, we then discuss existing defense techniques that achieve the objective and their effectiveness.
The order in which we discuss the elements is based on our insight that several approaches 
build on top of defenses for the preceding elements (our order also follows the rows in Table~\ref{tab:sok-table-compact}); for example, as resources encompass both tool artifacts and services, defenses for resources builds upon artifact defenses.

\begin{table*}[]
\caption{Supply chain defense techniques categorized by AStRA elements, security objectives and defense patterns. \fullcirc~denotes a technique that meets an objective and falls under a defense pattern. \halfcirc~denotes conditional/limited fulfillment. \text{\warning} denotes a potential conflict with other objectives. \checkmark denotes a defense technique that was developed specifically for the \swsc. Table~\ref{tab:sok-table2} in Appendix~\ref{app:taxonomy} contains our complete corpus.}
\footnotesize
\centering
%\begin{tabular}{|p{3.4cm}|P{1.8cm}|P{1.4cm}|P{1cm}|P{0.3cm}P{0.3cm}P{0.3cm}P{0.3cm}|P{0.2cm}P{0.2cm}|P{0.2cm}P{0.2cm}|P{0.2cm}P{0.2cm}|}
\begin{tabular}{|p{3.4cm}|P{1cm}|P{1.8cm}|P{1.4cm}|P{0.3cm}P{0.3cm}P{0.3cm}P{0.3cm}|lr|lr|lr|}
\hline
\multirow{3}{*}{\textbf{{AStRA Element}}} & \multirow{3}{*}{\makecell{\textbf{Supply} \\ \textbf{Chain} \\ \textbf{Specific?}}} & \multirow{3}{*}{\makecell{\textbf{Academic} \\ \textbf{Examples}}} & \multirow{3}{*}{\makecell{\textbf{Industry} \\ \textbf{Examples}}} & \multicolumn{4}{c|}{\multirow{3}{*}{\textbf{{Security Objective}}}} & \multicolumn{6}{c|}{\textbf{{Defense Pattern}}} \\
    &                   &                &   &     &     &     &     &     \multicolumn{2}{P{1.1cm}}{Detective} & \multicolumn{2}{P{1.2cm}}{Corrective} & \multicolumn{2}{P{1.3cm}|}{Preventative} \\
&                   &                 &  & & &  & & \rotatebox[origin=c]{90}{Transparency} & \rotatebox[origin=c]{90}{Integrity} & \rotatebox[origin=c]{90}{Graceful degr.} & \rotatebox[origin=c]{90}{Redundancy} & \rotatebox[origin=c]{90}{Limitation} & \rotatebox[origin=c]{90}{Correctness} \\
\hline
 \textbf{Principals} &                   &                 &  & \textbf{P1} & \textbf{P2} & \textbf{P3} & \textbf{P4} & 
& & & & & \\
ephemeral credentials        &     &   \cite{Newman2022}          &                   &   \fullcirc     &         &         &         &         &           &             &         &      \fullcirc                           &                    \\
secret scanning    &  & \cite{zhou_hey_2025}                 &                   &      &    \fullcirc     &         &         &    \fullcirc     &                               &                               &                               &                               &                               \\
credential transparency       &  & \cite{Newman2022}                &     \cite{ct}              &       &     \fullcirc    &         &         &    \fullcirc     &         \fullcirc                      &                               &                               &                      &             \\
role separation               &  &  \cite{Github-CI-Security}               &         \cite{gh-repo-roles, npm-roles}       &         &         &   \fullcirc      &      &         &                               &                  \fullcirc             &                               &                               &                               \\
multi-factor authentication   &      &             &   \cite{yubico}     &         \halfcirc      &         &  \fullcirc       &         &         &             \fullcirc                  &                               &                               &             \halfcirc                  &                               \\
separation of privilege &          &     &        \cite{cncf-best-practices}     &            &         &         &    \fullcirc     &         &                               &                               &                  \fullcirc             &                               &                               \\
threshold authorization     &    &    \cite{TorresArias2019, Newman2022}                  &          &         &         &     \fullcirc    &    \fullcirc     &        &                               &              \fullcirc                 &           \fullcirc                &                            &                        
\\[0.2em]
\hline
\textbf{\rule{0pt}{1.2em}Artifacts}           &      &             &                   & \textbf{A1} & \textbf{A2} & \textbf{A3} &     &     &                               &                               &                               &                               &                               \\
hashing        &    & \cite{vu_towards_2020} &                & \halfcirc       &         &       &         &         &        \fullcirc                         &         &                              &                                &                               \\
artifact signing      &   &     \cite{Newman2022}       &  \cite{git-gpg, github-commit-signing}  & \halfcirc            &       &        &         &    \fullcirc     &         &                               &                               &                               &                                                         \\
metadata signing      & \checkmark       & \cite{karthik_uptane_nodate}            &        \cite{docker-image-signing}         & \fullcirc       & \halfcirc       &        &        &   \halfcirc      &      \fullcirc                         &                               &                               &                             &                               \\
artifact transparency   & \checkmark     & \cite{Newman2022, merrill_speranza_2023} &       \cite{sigstore, scorecards}         & \fullcirc       & \halfcirc       &         &         &    \fullcirc     &         \fullcirc                      &                               &                               &                               &                             \\
ledger-based systems      &    &   \cite{contour_2018}      &              &    \fullcirc     &  \fullcirc        &       &         &    \fullcirc     &         \fullcirc                      &                               &                               &                             &                             \\
static analysis    &  & \cite{huang_vmud_2024, huang_donapi_nodate, ladisa_feasibility_2023} &   \cite{coverity}            &         &         & \halfcirc       &         &         &                               &                               &                             &                               &         \halfcirc                      \\
patch propagation    &   \checkmark &   \cite{dunlap_vfcfinder_2024, Luo2024, spider}      &              &    \text{\warning}  &        &    \fullcirc     &         &         &                               &                               &                             &                           &              \fullcirc                 \\
dynamic analysis    &  & \cite{HARP, he_rapidpatch_nodate, pantelaios_fv8_nodate} &               &       \text{\warning} &         & \halfcirc       &         &         &                               &                               &                              &              &            \fullcirc                   \\
ML-based analysis     &     &   \cite{Jiang_BinAI, Jia2024, jarhead}        &              &    \text{\warning}  &        &    \fullcirc    &         &         &                               &                               &                               &                           &                          \fullcirc      
\\[0.2em] 
\hline
\textbf{\rule{0pt}{1.2em}Resources}           &                   &      &          & \textbf{R1} & \textbf{R2} & \textbf{R3} &     &     &                               &                               &                               &                               &                               \\
mandatory access control   &        &          &     \cite{cncf-best-practices, apparmor}  &     \fullcirc    &         &         &         &         &                               &    \fullcirc                           &                               &                               &                               \\
filtering     &    & \cite{noauthor_natisand_nodate}         &       \cite{ebpf}        & \fullcirc       & \fullcirc       &    \halfcirc     &         &         &                              &                               &                               &             \fullcirc                  &                               \\
virtualization         &      &       \cite{sec-ci}            &    \cite{intel-vtx, amd-svm}           &         &    \fullcirc     &        \fullcirc  &       &         &                               &                               &                               &              \fullcirc                 &                               
\\[0.2em]  
\hline
\textbf{\rule{0pt}{1.2em}Steps}                  &                   &       &            & \textbf{S1} & \textbf{S2} & \textbf{S3} &     &     &                               &                               &                               &                               &                               \\
deterministic steps    & \checkmark  &      \cite{verifiability, dettrace, guix} &          &   \fullcirc     &         &         &         &         &                               &                               &                                 &                               &         \fullcirc                      \\
step transparency         & \checkmark      &  \cite{TorresArias2019}   &       \cite{in-toto, trivy}          &        & \fullcirc       &  \halfcirc   &   &   \halfcirc      &     \fullcirc    &                               &                             &                               &                                          \\
step consensus      & \checkmark         & \cite{chainiac, Policy-Transparency}          &       \cite{scitt}           &        & \fullcirc       &   \fullcirc      &         &   \fullcirc      &       \fullcirc                          &                            &                               &                            &                  \\

static analysis  &     &  \cite{wu_your_nodate,meng_identifying_2017}            &           &   \fullcirc      &         &         &         &         &                               &                               &                               &                               &              \halfcirc                 \\
ML-based analysis  &   & \cite{han_sigl_nodate}               &            &     \fullcirc    &         &         &         &         &                               &                               &                            &                               &           \fullcirc                    
\\[0.2em]
\hline
\textbf{\rule{0pt}{1.2em}Supply Chain Structure}    &               &     &              & \textbf{T1} & \textbf{T2} &     &     &     &                               &                               &                               &                               &                               \\
supply chain layout integrity & \checkmark & \cite{TorresArias2019}           &              & \fullcirc       &         &         &         &         &          \fullcirc                      &                              &                               &                               &                             \\
supply chain reproducibility &   \checkmark & \cite{verifiability}               &                 &         &    \fullcirc       &         &         &         &                               &                               &                               &                               &              \fullcirc                 
\\[0.2em] 
\hline
\end{tabular} \label{tab:sok-table-compact}
\end{table*}
\subsection{Defending Principals}
\label{sec:principals}

% Incidents like the initial stage of the SolarWinds hack~\cite{solarwinds-fireeye} and the Codecov breach~\cite{codecov} illustrate how attackers can compromise a key element of the \swsc by exploiting stolen credentials. These intrusions ultimately enabled the insertion of backdoors into the networks of thousands of companies and government agencies~\cite{reuters_solarwinds}. 

As in other systems, supply chain security begins with principal \textit{authentication} and \textit{authorization}. The defenses we observe are focused on protecting, detecting, and reducing authentication and authorization failures. 

To ground these concepts, we utilize the running example from Figure \ref{fig:topology}, where a Developer (Alice) acts as the primary Principal. Throughout this section, we illustrate how these defense patterns apply to Alice’s interactions with the build system to mitigate unauthorized access and tampering.

\Paragraph{P1 - Credential reuse limited.} Limiting the validity of a credential by time or use reduces how long an attacker can exploit it.

{\textbf{Ephemeral credentials}} constrain the window of compromise by limiting the time frame during which a credential is valid~\cite{cncf-best-practices}. 
The most common form used in the \swsc today are short-lived certificates~\cite{topalovic-certs},
digital certificates used to authenticate a remote party that have a short validity period, \ie they expire after
a short period of time (\eg a few minutes~\cite{sigstore-sec-model} to a few days~\cite{topalovic-certs}).
For instance, Sigstore's Fulcio~\cite{sigstore, fulcio} provides a framework for issuing short-lived X.509 certificates 
for code signing to curb the impact of compromised software distribution.
%In the case of the SolarWinds hack, this approach might have restricted how long attackers had access to the system.

\Paragraph{P2 - Credential compromise detected.} A key objective for securing principals is detecting whether credentials have been leaked, replaced, or misused.

{\textbf{Secret scanning}} analyzes resources searching for unintentionally
exposed credentials and cryptographic secrets~\cite{cncf-best-practices}.
Trufflehog~\cite{trufflehog}, for example, is capable of identifying different types of leaked secrets
on resources like git repositories and local file systems.
KeySentinel~\cite{zhou_hey_2025} improves secret scanning accuracy using machine learning and semantic analysis.

{\textbf{Credential transparency}} typically relies on a database backed by a Merkle tree
referred to as a \textit{transparency log}, which makes any changes to the contents of the log easily detectable.
Sigstore~\cite{sigstore} implements this approach for the \swsc 
by providing a specification and framework to leverage certificate transparency~\cite{ct} and audit the
validity of Fulcio-issued code signing certificates.
Some implementations~\cite{Newman2022} also preserve some security properties so long as there is one honest principal, 
providing additional reliability in the face of compromise (see \textbf{P4}).

%\noindent
%\begin{minipage}{\columnwidth}
%\vspace{6pt}
%  \fbox{%
%    \begin{minipage}{\columnwidth}
%      \textbf{Insight 1:} Detecting integrity and confidentiality problems with credentials is the first line of defense for reducing the risk to other \swsc elements.
%    \end{minipage}%
%  }
%  \vspace{6pt}
%\end{minipage}

\Paragraph{P3 - Credential authority limited.} Several techniques restrict the scope of a credential to increase the effort required to abuse a compromised principal.

{\textbf{Role separation}} limits each credential to a defined set of tasks by grouping principals according to their function in the supply chain~\cite{cncf-best-practices, Github-CI-Security,nist-800-204D}.
By minimizing shared authority, it reduces opportunities for interference and enforces least privilege.

% \msm{these roles don't really have much to do with the overall taxonomy, so I'm cutting this. maybe it fits in the appendix.}
% Common roles in open- and closed-source software development settings include project leads, maintainers, 
% and software developers and contributors, as well as project managers, product security engineers, system administrators, policy administrators, 
% software testers and project auditors, and customers (\ie product consumers). We derived this set of common \swsc roles from academic~\cite{ohm-backstabbers,ladisa-java,cappos-mirror-2008}
% and industry~\cite{cncf-best-practices,nist-ssdf} sources.

Platforms such as GitHub implement role-based repository permissions~\cite{gh-repo-roles}, and package managers such as npm apply similar role models~\cite{npm-roles}.

{\textbf{Multi-factor authentication (MFA)}} requires principals to authenticate using at least two credential categories. 
Following industry \swsc best practices~\cite{cncf-best-practices}, major package
managers~\cite{npm-2fa} and version control systems~\cite{gh-2fa} adopted two-factor authentication.

Most MFA integrations~\cite{gh-2fa, npm-2fa} in the supply chain leverage one-time passwords (OTPs) and
physical keys (\eg Yubikeys~\cite{yubico}) to limit credential reuse (\textbf{P1}).
%This could have made it harder for SolarWinds attackers to impersonate principals upon accessing the build system.

\Paragraph{P4 - Multiple principals required.} To limit the impact of a compromised or malicious principal, some defenses require that sensitive tasks be authorized or executed by multiple independent principals. 

{\textbf{Separation of principal privilege}} requires more than one principal to authorize a step, preventing any single actor from holding full authority. A common industry-recommended practice~\cite{cncf-best-practices} is two-person code review, in which at least two qualified reviewers must approve changes before they are merged. This reduces the risk of undetected issues and mitigates malicious contributions from malicious principals.

{\textbf{Threshold authorization}} divides the credential or signing capability for a step across $n$ principals, requiring a quorum to complete the operation. This not only raises the bar for unauthorized access, but also limits the authority of any single credential(\textbf{P3}).

The in-toto framework~\cite{TorresArias2019}, implements threshold models to maintain integrity even if some principals are compromised. Sigstore~\cite{Newman2022} similarly recommends threshold signing to reduce the likelihood that a compromised key can be used to release tampered artifacts.
%This approach might have narrowed the impact of the SOLARBURST attack~\cite{solarwinds-report}, or restrict the attackers' ability to access the build system.

\insightbox{Protecting principal authentication and authorization is the first line of defense for the \swsc. A combination of principal defense techniques seems necessary to significantly enhance overall supply chain security, with practical implementations available to meet most security objectives.}

% MSM: I think this is a good topic for the final discussion
% \noindent
% \begin{minipage}{\columnwidth}
% \vspace{6pt}
%   \fbox{%
%     \begin{minipage}{\columnwidth}
%       \textbf{Insight 1:} Layering and combination of principal defenses (e.g., ephemeral credentials with credential transparency) significantly enhance overall security but are underutilized in practice.
%     \end{minipage}%
%   }
% \end{minipage}

%\noindent
%\begin{minipage}{\columnwidth}
%\vspace{6pt}
%  \fbox{%
%    \begin{minipage}{\columnwidth}
%      \textbf{Insight 2:} Threshold authorization schemes show great potential but lack practical evaluations within the software supply chain domain.
%    \end{minipage}%
%  }
%\end{minipage}

%\noindent
%\begin{minipage}{\columnwidth}
%\vspace{3pt}
%%\colorbox{gray!10}{%
%\fcolorbox{black}{gray!10}{%
%
%    \begin{minipage}{0.97\columnwidth}
%  \textbf{Finding RQ2.1:} Principal defenses focus on credential compromise on all phases (pre- and post compromise)
%   \end{minipage}%
%  }
%  \vspace{3pt}
%\end{minipage}
\subsection{Defending  Artifacts}
\label{sec:artifacts}

Artifacts are the unit of exchange in the software supply chain. Attacks focus on surreptitiously modifying or replacing an authentic artifact with a compromised one.

In our running example Figure \ref{fig:topology}, the primary Artifacts are the Source Code and the final Docker Image. As the Source Code is transformed by the Build Step, defense patterns are applied to ensure that the resulting Docker Image accurately reflects the audited source and that its provenance is cryptographically verifiable by downstream consumers.

\Paragraph{A1 - Artifact changes detected.}
A core objective is detecting tampering with artifacts at rest, in transit, or in memory.

\textbf{{Hashing}} uses cryptographically strong hash functions (e.g., SHA-256) to detect bit-level changes. Because collisions are extremely unlikely, hashing enables artifact identification and content matching~\cite{nist-ssdf,cncf-best-practices,nist-800-204D} in settings such as source code~\cite{git-commit-hash} and package repositories~\cite{secure-apt}.
However, hashing alone only partially satisfies \textbf{A1}~\cite{git-attacks,legitimate}. If an attacker can replace a published artifact, they can typically replace its checksum as well. For instance, an attacker who can replace a published artifact on a website is likely able to also replace the checksum. In fact, given that these checksums are distributed through the same channel as the software artifact, any compromise of the channel subverts the integrity guarantees of hashing.

\textbf{{Artifact signing}} 
generates a digital signature over the contents of a software artifact (e.g., a package). A range of algorithms are used in practice: Sigstore~\cite{Newman2022} and Microsoft Authenticode~\cite{microsoft-authenticode} use x.509 certificates to sign OCI containers and device drivers, while Debian and Maven rely on GPG signing~\cite{secure-apt, apache-maven-gpg}. Version-controlled code, such as GitHub repositories, also commonly uses GPG signatures~\cite{git-gpg,github-commit-signing}. Because signature verification relies on trust metadata delivered through a separate channel, signing mitigates many of the weaknesses of hashing. Still, tampering with metadata can circumvent integrity guarantees even for signed artifacts~\cite{cappos-mirror-2008,git-attacks}, meaning artifact signing only partially fulfills \textbf{A1}.

\textbf{{Metadata signing}} extends these protections to metadata (e.g., file creation time), preventing attackers from manipulating metadata as an additional attack vector.
A widely deployed implementation is The Update Framework (TUF)~\cite{tuf}, which signs package repository metadata and uses role separation (see~\S\ref{sec:principals}) and publicly visible ``freshness claims''~\cite{tuf} to provide a degree of artifact non-equivocation (see \textbf{A2}).
Other variants of TUF exist for IoT and Automotive, notably, Uptane~\cite{karthik_uptane_nodate}. 

Similarly, signing Docker container manifests~\cite{docker-image-signing} transitively signs both the manifest and its associated metadata (e.g., layers, size, checksum). Combined with artifact signing, metadata signing can make unauthorized artifact changes fully detectable.

\Paragraph{A2 - Artifact non-equivocation.} 
This objective ensures that all principals observe the same artifact produced and consumed by a step. 

\textbf{{Artifact transparency}} uses transparency logs to provide a verifiable 
history of an artifact's changes over time.
This makes artifact tampering and surreptitious or compromised releases detectable (\textbf{A1}),
but the degree of transparency depends on the log's deployment (local or public).
Because credentials are a special case of artifacts, this mechanism naturally generalizes credential transparency (see~\S\ref{sec:principals}).

Examples include Scorecards~\cite{scorecards}, which analyzes source repositories for security-related practices and assigns a security posture score~\cite{cncf-best-practices}; Firefox’s Binary Transparency~\cite{ff-ct1}, which published signed binaries in a public log; and Sigstore’s Rekor log~\cite{rekor}, which records a wide range of artifacts, including
signatures themselves, in-toto attestations~\cite{in-toto-attestation}, 
software bills of materials, and other arbitrary binary blobs.
Speranza~\cite{merrill_speranza_2023} takes a privacy-preserving approach by using anonymous credentials and zero-knowledge proofs so that only authorized signers can produce valid signatures without revealing their identities.
%Similar solutions also exist in grey literature and industry to provide similar mechanisms.

However, transparency logs are vulnerable to fork attacks unless supplemented with additional monitoring mechanisms~\cite{rekor-monitor}. As a result, artifact transparency alone only partially satisfies \textbf{A2}.

\textbf{{Ledger-backed systems}} strengthen non-equivocation guarantees by using distributed ledgers (DLs) to make tampering and omissions \textit{globally} visible. The ledger’s append-only structure and consensus protocol provide strong tamper-detection properties (\textbf{A1}).
Contour~\cite{contour_2018}, for example, leverages Bitcoin's ledger to help users detect these attacks. 

%\noindent
%\begin{minipage}{\columnwidth}
%\vspace{6pt}
%  \fbox{%
%    \begin{minipage}{\columnwidth}
%      \textbf{Insight 3:} Ledger-backed systems is promising, provide availability and privacy properties~\cite{scitt} in addition to (\textbf{A1, A2}), but face significant usability and adoption challenges, necessitating more research into performance optimizations and user-centered evaluations.
%    \end{minipage}%
%  }
%  \vspace{6pt}
%\end{minipage}

\Paragraph{A3 - Expected artifact behavior enforced.} Defenses under this objective ensure that artifacts behave as intended. %Techniques generally either characterize normal behavior to distinguish it from malicious activity or directly identify abnormal behavior.

\textbf{{Static analysis}} is used to
identify suspicious components such as installed packages (e.g., by detecting a vulnerable dependency), or vulnerabilities that may be introduced transitively (e.g., a statically compiled vulnerable
library~\cite{nist-ssdf,cncf-best-practices}). 
At the core of these approaches lies a mechanism for evaluating artifact representations, such as an abstract syntax tree (AST) or binary code (\eg~\cite{coverity}).
%Second, an enforcement mechanism that can be integrated into a supply chain element (e.g., within a package manager resource, or a CI/CD step).

Some systems~\cite{krohmer-scored, noauthor_argus_nodate} use static taint analysis to trace flows through artifacts such as PHP web applications and GitHub Actions workflows, identifying risky execution paths and misconfigurations. Others detect abnormal code behavior in specific language code (for example,~\cite{ladisa-java, jarhead} focus on Java). CodeQL-based analysis~\cite{gobbi_poster_2023} identifies malicious JavaScript patterns (e.g., suspicious API usage).

Static comparison of source code and corresponding published artifacts~\cite{vu_towards_2020} can reveal unauthorized modifications introduced during build or release (\textbf{A1}).
Recent work also explores vulnerability detection via feature classification of reused code components~\cite{woo_v1scan_2023}, and static analysis is increasingly embedded into DevSecOps workflows for automated risk enforcement~\cite{noauthor_product_nodate}.

\textbf{{Patch propagation}} builds on static analysis to generate and apply source code patches that propagate security patches to forked codebases~\cite{spider}.
Similarly, VFCFinder~\cite{dunlap_vfcfinder_2024} and SPatch~\cite{Luo2024} use symbolic code execution and natural language processing, respectively, to identify, generate and apply patches to source code.
Unlike direct static analysis techniques, patch propagation not only detects anomalous behavior but also aims to correct it.

\textbf{{Dynamic analysis}} %evaluates an artifact's behavior during its execution.  The execution may be emulated or performed in a controlled environment to reduce the influence of external factors in the analysis. In the context of software supply chains, this approach is used to detect maliciously behaving artifacts before their utilization as a resource (see~\S\ref{sec:resources}). The goal of this approach is twofold: to avoid maliciously-injected malicious behavior, and to reduce the number of vulnerabilities introduced as dependencies in the software supply chain. 
FieldFuzz~\cite{bytes_fieldfuzz_2023}, 5G-Spector~\cite{wen_5g-spector_2024}, and FV8~\cite{pantelaios_fv8_nodate} dynamically analyze artifacts ranging from automation runtimes and cellular infrastructure to JavaScript packages, revealing threats in third-party components. Hybrid analyses combine static and dynamic signals: DECAF~\cite{christensen_decaf_2020} removes unused firmware while preserving runtime correctness; RapidPatch validates automatically generated firmware hot-patches~\cite{he_rapidpatch_nodate}; DONAPI~\cite{huang_donapi_nodate} detects malicious npm packages by integrating static graphs with runtime traces; and~\cite{ntousakis_demo_2021} identifies subtle API misuse in third-party libraries.

\textbf{{Machine learning (ML) based analysis}} is an emerging line of work. Jarhead~\cite{jarhead} combines static analysis and ML to classify Java packages as benign or malicious. Ladisa et al.~\cite{ladisa_feasibility_2023} and Latch~\cite{wyss_wolf_2022} apply tree-based models to detect malicious npm and Python packages.

Active Learning and Regeneration (ALR) systems such as HARP~\cite{HARP} and Up2Dep~\cite{nguyen_up2dep_2020} detect abnormal application behavior—such as suspicious file system changes or global state manipulation—by modeling expected functionality in a controlled environment to avoid learning malicious patterns. These models can regenerate vulnerability-free versions of libraries in JavaScript and C/C++.
Recent work~\cite{noauthor_poster_nodate} also uses large language models to learn semantic patterns that distinguish benign from malicious source code, even across varied coding styles and obfuscation techniques.
Although dynamic and ML-based techniques can effectively enforce expected behavior, their ability to rewrite or regenerate code or binaries may conflict with the detection of artifact changes (\textbf{A1}).

%\noindent
%\begin{minipage}{\columnwidth}
%\vspace{6pt}
%  \fbox{%
%    \begin{minipage}{\columnwidth}
%      \textbf{Insight 4:} Dynamic analysis methods can inadvertently conflict with other artifact objectives, suggesting the need for careful cross-objective evaluation when deploying defenses.
%    \end{minipage}%
%  }
%\end{minipage}
%\noindent
%\begin{minipage}{\columnwidth}
%\vspace{3pt}
%%\colorbox{gray!10}{%
%\fcolorbox{black}{gray!10}{%
%
%    \begin{minipage}{0.97\columnwidth}
%  \textbf{Finding RQ2.2:} Most artifact defenses provide either integrity or correctness through various means.
%   \end{minipage}%
%  }
%  \vspace{3pt}
%\end{minipage}
\subsection{Defending Resources}
\label{sec:resources} As described in ~\S\ref{sec:model}, resources are themselves software components, \ie they are created
through their own software supply chains. This means that many security objectives that apply to resources
may be met by artifact defenses (see~\S\ref{sec:artifacts}).
At the same time, because resources are used to carry out (parts of) a particular supply chain step,
we identify additional security objectives specific to resources, which pertain to mitigating 
execution-time risks.

For a resource to operate, it typically requires auxiliary resources and artifacts,
which we generally refer to as the \textit{execution environment}.
Depending on the type of resource, it execution environment may be defined at various 
levels of granularity ranging from an intra-process unit (\eg thread or domain), to an application, to a virtualized environment (\eg container or VM), to an entire system running on a hardware platform.
Thus, to reduce risks in resources we must protect different aspects of execution environments.

In our running example Figure \ref{fig:topology}, the primary Resource is the GitHub Actions Runner. By applying defense patterns to this resource, we ensure that the environment used to execute the 'Compile' step is authenticated and hasn't been tampered with before the process begins.

The Log4j vulnerability highlights the supply chain risks associated with widespread use of open-source dependencies as resources. %The open-source logging utility hosted by the Apache Foundation~\cite{log4j, apache-foundation}, logs messages within software and has the ability to communicate with other services on a system. 
In December 2021, a vulnerability was discovered in the log4j utility, which that enabled remote attackers to inject and remotely execute code through log messages. After its discovery, millions of attempted exploits, many of which turned into successful denial-of-service (DoS) attacks~\cite{uberti_log4j_2021}. 

%%% TEXT

%This includes artifacts being used as tools (e.g., a compiler), or inputs (e.g., as source code).
%\vspace{-.5ex} %MSM counteracts the extra space above the top \Paragraph
\Paragraph{R1 - Resource privilege limited:} Limits the operating privileges a 
resource is granted to only those needed to carry out a step \textit{within}
its execution environment. This objective is equivalent to \textbf{P3}
aiming to reduce risk within the context of an authorized step.

%on that resource, such that strict security controls are not compulsory. One major drawback of this technique is minimal protection from malware running as privileged normal user. With that, the superuser account represents a single point of failure, an attacker can compromise the whole system by taking over a process under his privileges, since he is the owner of the system. 
%\cite{https://dl.acm.org/doi/fullHtml/10.5555/1149826.1149839}

{\textbf{Mandatory Access Control (MAC)}} enforces a non-discretionary,
global access policy which enforces multiple levels of security for operations resources perform~\cite{least-priv}.
As such, MAC can be used to meet objective (\textbf{R1}) as a compromised resource
can be prevented from performing an unauthorized operation by default.

One way these defenses may help reduce \swsc risk is by running a kernel-level MAC system
like AppArmor~\cite{apparmor} or SELinux~\cite{selinux} to control which operations (\eg \texttt{read()}, \texttt{mount()}) 
a CI/CD or build resources may perform on artifacts~\cite{cncf-best-practices}.
% MSM: this is a very strong claim, and details that are too in the weeds
%no one single operation that can be used to compromise the whole system, while the superuser account is used for maintaining the global security policy ~\cite{https://dl.acm.org/doi/fullHtml/10.5555/1149826.1149839}.  However, setting security policies can be hard and complicated because of the needed knowledge of the intended behavior of every application/user on the system.
%One  of MAC is AppArmor security model, which binds access control attributes to programs rather than to users, with the intention of the biggest attack vector on most systems is application vulnerabilities ~\cite{https://wiki.ubuntu.com/AppArmor }. So, each program may have a profile loaded in the kernel while booting. A profile can be either in enforcement or complain mode, based on that, the policy can be either enforced, or just reporting policy violation attempts. 
%AppArmor functionality is integrated in the mainline Linux kernel from 2.6.36 onwards, but is mainly used in SUSE and Ubuntu.

%- policy writing , hard to configure and maintain
\comment{
\Paragraph{Role-Based Access Control (RBAC) } 
 This approach controls the access to the resources based on assigned users roles in the system. 
 A role describes the user function, responsibility, such as: owner, contributor, or even VM reader. This model can be used to enforce DAC, and MAC on any role group. Security Enhanced Linux (SELinux) is a good example of RBAC that implements MAC for role groups. 
 SElinux is included in the mainline Linux kernel from 2.6, but primarily used on Redhat/Fedora systems

%not sure if it is used in the ssc context
\Paragraph{Rule-Based Access Control } 
In this mechanism, the OS grants permissions based on set of predefined rules and policies, such rules may be role based, or may depend on parameters like access time or location.

- Flexible but  lots of admin work to be monitor, change the rules

{\textbf{Other access control approaches.}}
For brevity, we focus on the two main models that meet resource security objectives.
Specific access control strategies mostly vary by granularity and types of policy rules,
but they may be used to enforce DAC or MAC models. Examples include Role-Based Access Control (RBAC)~\cite{rbac},
Attribute Based Access Control (ABAC)~\cite{abac}, and History-Based Access Control~\cite{hbac}. In the context of software supply chains, Artemis~\cite{Moore2023} applies RBAC at the package level by allowing repositories to explicitly scope developer authority, preventing unauthorized publication or hijacking of packages.
}
\Paragraph{R2 - Environment interfaces restricted:} Restricts the access points to
and from a resource's execution environment.

{\textbf{Filtering}} techniques intercept accesses at a particular
environment interface controlling the level of interactions between internal
and external resources. 
HODOR ~\cite{noauthor_hodor_nodate} applies combined static-dynamic call graph analysis to enforce thread-specific seccomp syscall filtering in Node.js, limiting resource privileges \textbf{R1} and restricting environment interfaces \textbf{R2} , with partial enforcement of execution boundaries (R3).
Defenses such as eBPF~\cite{ebpf} provide programmable filtering of a
variety of kernel-level interfaces (\eg network, system calls), which also enables
enforcement of MAC policies for specific resources running on a system (\textbf{R1}, \textbf{R2}). While eBPF~\cite{ebpf} executes within an isolated environment within the kernel, we do not consider this sufficient to meet \textbf{R3} because boundaries are only enforced around the defense mechanism itself.
NatiSand~\cite{noauthor_natisand_nodate} similarly applies eBPF, and other techniques to sandbox native modules in JavaScript runtimes, enforcing syscall-level policies to restrict untrusted dependencies.

\Paragraph{R3 - Environment boundaries enforced:}
Explicitly defines and enforces the boundaries of a resource's execution environment. 
This objective aims to prevent any interference between different resource
environments, even in the face of any fault or compromise.

% \Paragraph{R2 - Scope of single resource function limited:} Narrows the scope of each function performed by a resource on an artifact 
% or another resource. This property is achieved by controlling the level of authority of an individual resource function at a finer granularity, also known as \textit{separation of privilege} in the literature~\cite{least-priv}.

{\textbf{Virtualization}} creates an abstract execution environment in which resources share 
common underlying hardware and software components without needing to be aware of other environments~\cite{nist-isolation}.
Software virtualization techniques establish boundaries based on specific common properties or conditions shared by resources that 
are typically enforced by privileged software (\textbf{R3}).  % while enforcing separation between these processes as well as constraints within on resource use within these process groups. 
%This reduces the startup time and allows more virtual instances to run on the same hardware. 
%As such, kernel-based isolation techniques are a fundamental technique used in container-based execution environments.
Hardware virtualization approaches (\eg~\cite{intel-vtx,amd-svm}), on the other hand, are designed to enforce strong boundaries and controlled I/O 
interfaces using dedicated CPU features (\textbf{R2, R3}). 
Virtual machines are very commonly used in the \swsc to implement strongly isolated execution environments
for resources such as CI/CD runners~\cite{sec-ci}.
%virtualization, for instance, multiple virtual instances of a device run on a single physical hardware resource. Each instance or virtual machine executes with its own copy of an operating system (OS), libraries, dedicated resources, and applications. This overhead results in high startup time and supports running less number of VMs on the same hardware.

\comment{
\Paragraph{Sandboxing} is \textit{application isolation}  technique   that  control execution environment that prevents potentially malicious software, such as mobile code, from accessing any system resources except those for which the software is authorized\cite{nist-isolation}.
}

%- mention eBPF
\comment{
\subsubsection{Preserving Minimum Security Properties}
Survivability \cite{} is the ability for a system to function correctly while under attack or partial compromise.

\Paragraph{The Update Framework (TUF)}improves the software update system key compromise survivability using responsibility separation and delegation, multi-signature trust with threshold signatures and multiple roles, and explicit and implicit key revocation. 
}

%diplomat
%\noindent
%\begin{minipage}{\columnwidth}
%\vspace{6pt}
%  \fbox{%
%    \begin{minipage}{\columnwidth}
%      \textbf{Insight 5:} Industry best practices heavily rely on  resource defense mechanisms ~\cite{cncf-best-practices,nist-ssdf,slsa}, highlighting a need for further academic evaluations to measure their practical effectiveness and limitations.

 %   \end{minipage}%
  %}
%\end{minipage}

%\noindent
%\begin{minipage}{\columnwidth}
%\vspace{6pt}
%  \fbox{%
%    \begin{minipage}{\columnwidth}
%      \textbf{Insight 6:}  Defenses for resources have mature underlying techniques (MAC, DAC, filtering, virtualization) but still lack sufficient research in the context of software supply chains.

%    \end{minipage}%
%  }
%\end{minipage}
%\noindent
%\begin{minipage}{\columnwidth}
%\vspace{3pt}
%%\colorbox{gray!10}{%
%\fcolorbox{black}{gray!10}{%
%
%    \begin{minipage}{0.97\columnwidth}
%  \textbf{Finding RQ2.3:} Resource defenses focus on privilege separation in the execution environment.
%   \end{minipage}%
%  }
%  \vspace{3pt}
%\end{minipage}
\subsection{Defending Steps}
\label{sec:steps}

Steps represent software development actions or operations that principals carry out on 
artifacts by use of resources.
%Steps are the center of the AStRA model, and thus, reducing their risks is paramount.
Generally, step risks depend on whether risks to principals, resources and artifacts
in a step have been addressed. 
For example, compromising a principal who is authorized to carry out a step leads to
a compromised step.

Yet, even with reduced risks to principals, artifacts and resources, steps may still face
risks related to the invocation or operation of the step.% \textbf{MSM: Integrate this next paragraph here!}
% A step typically consumes and produces software artifacts; A \textit{transformation} step produces one or 
% more software artifacts from consumed input \textit{materials} (\eg a compilation step), while an 
% \textit{informational} step provides auxiliary information about artifacts or other supply chain components;
% examples include operations such as static analysis (artifact level), or an ISO process audit (structure of the software supply chain)~\cite{ite-10}.

%\vspace{-.5ex} %MSM counteracts the extra space above the top \Paragraph
%Several mechanisms exist to reduce risks to carrying out steps.
In our running example Figure \ref{fig:topology}, the primary Steps are 'Compile' and 'Containerize'. By applying defense patterns to these steps, we ensure that the transformation from Source Code to Docker Image is isolated from external interference and that the process itself produces verifiable evidence of its execution.

While most of the approaches we evaluate target build and packaging steps, 
their design principles and properties can be applied to other steps in the \swsc.
These mechanisms fall within two major categories: Preventative and detection approaches.

\Paragraph{S1 - Expected step behavior enforced:} The expected step behavior is well-defined
and cannot be altered once triggered. Ensuring correct step behavior may be achieved
by the correctness of the resource(s) used to carry out a step.

{\textbf{Deterministic steps}} follow patterns that prevent sources
of entropy from influencing a steps' behavior or outcome (\textbf{S1}).
For instance, Court\`es~\ea~\cite{guix}, describes how GUIX ensures correctness via 
source-to-binary mapping. %and granular version control allows for step correctness, transparency and non-equivocation.

Navarro-Leija et al. proposed DetTrace~\cite{dettrace}, a reproducible container abstraction that
enables reproducible software builds, among other use cases. 
To this end, DetTrace controls the influence of a system's behavior, such as naturally noisy system calls
(\eg \texttt{clock\_gettime}) that would introduce nondeterminism into a step (\textbf{S1}).

{\textbf{ Static analysis}}
%\msm{this whole para below belongs under steps}
Other approaches apply static techniques to model authorship fingerprints in compiled binaries~\cite{meng_identifying_2017}, match recurring vulnerability fixes via semantic equivalence~\cite{huang_vmud_2024}. Static analysis has also be applied detecting missing verification steps in firmware update mechanisms~\cite{wu_your_nodate}.
Since static analysis techniques tend to be domain-specific and require a separate enforcement mechanism, however, this approach only partially enforces expected artifact behavior.
% \msm{Mandated tools/libraries/mechanisms, from Wurster~\ea:} ``This works well for large companies with leverage (e.g. forcing
% third-party developers to use analysis tools, specific languages, specific compiler flags, and other elements). Mandates can also come from management within a company (e.g., forcing all developers to run a static analysis tool). Clients, system vendors, and others can also mandate the use of security tools.''

\Paragraph{S2 - Step operation recorded:} The identities of principals, artifacts used 
and produced, and resources used, are recorded. Further, a step provides faithful 
information about its operation.

{\textbf{Step transparency}} provides information about any aspect of a step's operation to provide 
visibility into all aspects of a step (\textbf{S2}).
A prominent type of step metadata are Software Bills of Materials~\cite{sbom}
for their use in \swsc compliance~\cite{eo-memo}. Their main goal, while not authenticated, 
is to convey information about artifact composition, software ``pedigree'', and quality, 
complementing other operational metadata such as SLSA Provenance~\cite{slsa}.

One of the earliest proposals to capture step operational metadata was in-toto~\cite{TorresArias2019}
via a primitive referred to as Links. in-toto Links are \textit{authenticated} documents describing
the input and output artifacts to a step, the command executed during a step, and any step
byproducts such as return value or error messages. Using principal and artifact defense mechanisms
allows in-toto to also detect tampering with any aspect of the step (\textbf{S3}).

\Paragraph{S3 - Interference with step detected:} Detects attempts to tamper with or 
otherwise interfere with the expected behavior of a step. Detecting attempts to present different
versions of the same step to different principals provide the stronger global non-equivocation property.
\comment {
\Paragraph{S4 - Step containment:} A step's behavior must minimize the impact of 
compromise on the remaining \model to prevent it from affecting subsequent steps.
%\msm{We might want to use the ``containment'' language here, since this one is really about fault containment.} 
\Eman{I'm wondering if this fits more in the topology section? its a subset of T3}
}

{\textbf{Step consensus}} relies on the repeatability and/or visibility of a step to enable multiple parties to determine jointly if a step has been tampered with
(\textbf{S3}). That is, if different parties obtain different metadata or results for a given
step, then the step is assumed to have been tampered with.

Approaches such as the IETF Supply Chain Integrity Transparency and Trust (SCITT~\cite{scitt}) framework
and Ferraiuolo~\ea~\cite{Policy-Transparency} provide global (\textbf{S2}) and (\textbf{S3}) properties
using mechanisms like distributed ledgers and transparency logs to expose policy decisions as part of the
metadata recording and publication process.
% For instance, Succinctly put, they propose a mechanism that ensure \swsc policy decisions are carried out at the time of admission into a transparency mechanism (i.e., a ct-log or a confidential-compute-backed ledger~\cite{azure-confidential-ledger}). 

%\msm{This one might fit better in the topo section}\Eman{I think the same}

% \msm{This is all seems artifact-related...}
% Al-Bassam et al.~\cite{contour_2018} designed Contour, a binary transparency system, leveraging Bitcoin blockchain to proactively and efficiently prevent users from installing malicious software. The system guarantees transparency, privacy, and availability of software package binaries in a way that breaking its integrity costs millions of dollars even for attackers who can carry out a man in the middle attack. Contour was tested to audit software binaries in the Debian software repository and showed easy deployment on the ecosystem with relatively low overhead to current infrastructure, and with no changes or coordination requirements for any participant.

Nikitin~\ea proposed CHAINIAC~\cite{chainiac}, for instance, provides strong tamper-detection guarantees
for software updates through a novel data structure called a skipchain (\ie the merging of a skiplist and a blockchain). 
CHAINIAC detects step tampering in two stages. First, a group of build services re-runs 
a particular build step. Then, they validate the resulting binary, and only commit it into the skipchain
if the builders reach consensus over the binary.

As CHAINIAC skipchains contain a cryptographically verifiable
log of all released binaries, the approach also helps different parties ensure that they all
have the same view of the software updates over time (the stronger non-equivocation property of \textbf{S3}).

%\msm{static analysis/ML based analysis missing? per table 2}

\comment{
\begin{table}[t]
\center
\footnotesize
\caption{Step defense approaches mapped to the security properties they provide.}
\label{tab:step-properties}
\begin{tabular}{l c *{3}c}
    \toprule
    \textbf{Technique} & \textbf{Example} & 
    \textbf{S1} & \textbf{S2} & 
    \textbf{S3} \\
    \midrule
    Formal verification$^\dagger$ & Verifiable Compilers~\cite{dafny} & \fullcirc & - & -  \\
    Deterministic steps$^\dagger$ & DetTrace~\cite{dettrace} & \fullcirc & - & -  \\
    Step transparency & SBOMs~\cite{spdx} & - & \fullcirc  \\
    Auth. Step Transparency$^\dagger$ & in-toto~\cite{TorresArias2019} & - & \fullcirc & \fullcirc \\
    Step Consensus$^\dagger$ & CHAINIAC~\cite{chainiac} & - & \halfcirc & \fullcirc  \\
\bottomrule
\end{tabular}

% MSM: this space is needed to ensure the table legend is rendered below the table
\medskip
\fullcirc~= provides property; \halfcirc~= conditionally provides property; -~= property not provided; $\dagger$~=has academic publication
\end{table}
}

%\noindent
%\begin{minipage}{\columnwidth}
%\vspace{6pt}
%  \fbox{%
%    \begin{minipage}{\columnwidth}
%      \textbf{Insight 7:} Formal verification methods have proven successful in other domains but are largely unexplored for ensuring the correctness of software supply chain steps    

%    \end{minipage}%
%  }
%\end{minipage}

%\noindent
%\begin{minipage}{\columnwidth}
%\vspace{6pt}
%  \fbox{%
%    \begin{minipage}{\columnwidth}
%      \textbf{Insight 8:} Consensus-based and containment defenses for supply chain steps are theoretically well-developed; however, future research should focus on evaluating their practical applicability, usability, and adoption costs.   
%    \end{minipage}%
%  }
%\end{minipage}

%\noindent
%\begin{minipage}{\columnwidth}
%\vspace{3pt}
%%\colorbox{gray!10}{%
%\fcolorbox{black}{gray!10}{%
%
%    \begin{minipage}{0.97\columnwidth}
%  \textbf{Finding RQ2.4:} The majority of step defenses focus on mechanisms to record step information and detect misbehavior.
%   \end{minipage}%
%  }
%  \vspace{3pt}
%\end{minipage}
\subsection{Defending Supply Chain DAG Structure}
\label{sec:topology}

%%% TABLE
% \comment{
% \begin{table}[t]
% \center
% \footnotesize
% \caption{\swsc topology defense approaches mapped to the security properties they provide.}
% \label{tab:topo}
% \begin{tabular}{l c *{5}c}
%     \toprule
%     \textbf{Technique} & \textbf{Example} & \multicolumn{4}{c}{\textbf{Security}} \\
%     & &
%     %risks  
%     \textbf{T1} & \textbf{T2} & \textbf{T3} & \textbf{T4} & \textbf{T5} 
%     \\
%     \midrule
%     supply chain layout integrity$^\dagger$ & in-toto layouts~\cite{TorresArias2019} & \fullcirc & - & - & \fullcirc &-\\
%     resource replication &  Mirrors~\cite{apt-mirror} & - & \fullcirc & - & -&-\\
%     dependency reduction & Bootstrappable builds~\cite{bootstrappable-builds} & - & - & \fullcirc & -&- \\
%     %principal equivalence & context-based access control~\cite{cbac} & - & - & - & \fullcirc \\
%     supply chain reproducibility & Reproducible builds~\cite{reproducible-builds} & - & - & - &\fullcirc& \fullcirc \\
% \bottomrule
% \end{tabular}

% % MSM: this space is needed to ensure the table legend is rendered below the table
% \medskip
% \fullcirc~= provides property; \halfcirc~= conditionally provides property; -~= property not provided; $\dagger$~=has academic publication
% \end{table}
% }
%%% TEXT

The left-pad incident~\cite{leftpad}, in which the package's removal
from the npm repository broke downstream build steps that use it as a dependency,
exemplifies the risks propagated via the relationships between supply chain elements.
While previous sections cover individual supply chain elements,
a small number of \swsc defenses focuses on maintaining and securing their
\textit{transitive} relationships.

\Paragraph{T1 - DAG changes detected.} To validate the integrity of relationships between elements,
changes to the structure of a supply chain DAG, \ie additions, removals or reordering of 
individual elements, can be made detectable.

{\textbf{Supply chain layout integrity}} enables principals to detect whether all steps were performed in the 
expected order, whether they were carried out by the expected principals, and if artifacts were modified between steps.
This approach first introduced in Torres-Arias~\ea~\cite{TorresArias2019} reduces the risks of added, removed, or reordered steps.

Using a primitive called layouts, in-toto~\cite{TorresArias2019} implements this approach by requiring project owners or maintainers 
(a principal role) to pre-define the set of steps to be performed, the principals authorized to perform each step, and the expected consumed and produced materials for each step, a priori.
Given an in-toto layout document authenticated by the expected project owner, and step metadata is available (see~\S\ref{sec:steps}), 
interested principals can validate that an artifact was produced following the expected layout.

Beyond package dependency graphs, few mechanisms exist today to analyze the security and integrity of supply chain DAGs.
More human-centered research is needed to understand how practitioners configure software supply chains DAGs, including all their elements, enabling a deeper understanding of how structural risks surface, and efforts to establish standard, robust DAG structures.

\Paragraph{T2 - Consistent results.} Checking whether a set of steps performed on artifacts in a specific order always produces consistent results can confirm the correctness of a DAG. Techniques that check for exact matches in final artifacts provide the strongest form of consistency.

{\textbf{Supply chain reproducibility}}, first proposed in the 1990s, seeks to detect whether a particular sequence of steps
always produces the same results. In practice, reproducibility relies on the chaining of techniques for individual steps like deterministic builds (\S\ref{sec:steps}), which ensure that a particular step, such as build, always produces the same bit-for-bit result.
By enabling downstream principals to repeat a sequence of upstream steps, upstream producers enable their consumers to also detect
changes in the supply chain DAG for their product artifact.

%Supply chain reproducibility is more widely explored in practice~\cite{citespls} than in academia~\cite{reproducile_builds_usability}. 
As discussed in~\cite{reproducile_builds_usability}, most prior work has focused on addressing the technical challenges of reproducibility; more human-centered studies on the daily challenges of achieving and maintaining reproducibility are needed to improve adoption and lead to consistent supply chain DAGs.

% \Paragraph{Discussion}
% Using topology information from different data sources, it is possible to shed light on the correlation between different supply chain practices and their security outcomes to provide empirical models to harden the software ecosystem.
% \msm{This previous sentence needs some tie-in to the text or reference, and I'm not sure what it's trying to say. I have moved the rest of this discussion to the future research section.}

%\noindent
%\begin{minipage}{\columnwidth}
%\vspace{6pt}
%  \fbox{%
%    \begin{minipage}{\columnwidth}
%      \textbf{Insight 9:} Very few defenses target the transitive relationships between elements in the supply chain. Of the few techniques that are available, all assume that expected or correct relationships have been established a priori.
%    \end{minipage}%
%  }
%\end{minipage}
\insightbox{We identified 13 security objectives for the core \swsc elements (principals, artifacts, resources steps), and two extra objectives at the DAG structure level.}
\subsection{Prevalent Defense Patterns}
\label{sec:defense-patterns}

Through our systematized security objectives and defense techniques,
we observe prevalent strategies to how software supply chain defenses address
risk \textit{across} different elements.
Thus, we derive six defense patterns, which we group into three
general security control types~\cite{wright_control_types2008}.

More formally, \textbf{a defense pattern describes
the scope, context and outcome of a defense technique}.
That is, the security objective and target of a defense technique (its scope)
commonly dictate when and where in the supply chain it must be implemented
(its context).
A defense technique's scope and context, in turn, affect its outcome,
\ie the time when a particular supply chain risk is identified
and the extent to which it is mitigated.

We map our objectives in the taxonomy to the defense patterns in the last column group of table~\ref{tab:sok-table-compact}.

%Our taxonomy shows that there is no one defense technique that addresses
%the full range of security objectives across the software supply chain.

\subsubsection{Detective Defenses}
\label{sec:defense-patterns:detective}

Approaches in this category are \textit{reactive} as their outcome is that
supply chain participants retroactively determine whether a threat occurred
in the past.
Such defense techniques are typically separated into two components: 1) a
monitor or observer that collects security properties or behaviors
of a supply chain element, and 2) a validator that is responsible for
examining or verifying the collected supply chain information as
at a later point in time.
In the event a threat has occurred, the information collected by the monitor
can serve as evidence to \textit{detect} the problem and alert supply chain
participants.

\textbf{{Transparency}} defenses have the goal of making any property
about a supply chain element visible and verifiable by \swsc principals.

All defenses that detect principal credential compromise (\textbf{P2}) offer
transparency because they reveal confidentiality or integrity issues.
Similarly, the observed artifact equivocation (\textbf{A2}) defenses, which
provide artifact transparency, aim for supply chain elements to have the 
ability to recognized when their views of an artifact diverge. 
Step transparency defenses record step operation information (\textbf{S2}),
making \swsc processes visible.

\textbf{{Integrity}} techniques seek to enable \swsc principals to verify
that a supply chain element is authentic, and compare two supply chain elements
are equivalent.

Credential transparency not only provides enhanced auditability of
principal credential usage, but also credential integrity by making
mis-issued or tampered credentials detectable (\textbf{P2}).
All techniques that detect changes in an artifact (\textbf{A1}) rely on cryptographic mechanisms to check artifact integrity, i.e.,  if it meets an expectation.

Similar to credential transparency, step transparency is a defense that
covers both transparency and integrity goals by making interference
with a supply chain step verifiably detectable (\textbf{S3})
Today's defenses, on the other hand, that aim to make \swsc DAG changes
detectable (\textbf{T1}) largely focus on providing integrity properties.

\subsubsection{Corrective Defenses}
\label{sec:defense-patterns:corrective}

These defenses are \textit{active} as their outcome is to reduce the impact
of actively occurring supply chain problems,
including availability attacks.
That is, even in the face of an attack or vulnerability,
such defenses \textit{correct} the supply chain element to a known good
or stable state.

Like preventative defenses, corrective defenses are in place a priori,
but they are triggered after an incident takes place.

\textbf{{Graceful degradation}} defenses have the objective to
maintain a baseline level of security in the \swsc, even when a compromise
has occurred.

For principals, defenses that limit credential authority (\textbf{P3})
ensures that supply chain steps the a compromised credential is not
authorized to access remain inaccessible.
Similarly, mandatory access control in the \swsc aims to maintain a
baseline level of access restrictions to a resource (\textbf{R1}).

\textbf{{Redundancy}} techniques generally rely on a fallback mechanism
to restore a particular supply chain element to normal function in the face
of a compromise.
The basis of all principal defenses requiring multiple principals to
perform an authorization (\textbf{P4}) is to have sufficient uncompromised
principals available in the event of an attack for a \swsc step
to be carried out as expected.

\subsubsection{Preventative Defenses}
\label{sec:defense-patterns:preventative}

These defenses are \textit{proactive} as their outcome is to reduce the impact or prevent a specific threat
from occurring altogether.
Such techniques are applied before their target supply chain element is deployed in anticipation of potential risks, and thus they techniques follow a \textit{preventative} pattern.

\textbf{{Limitation}} defenses restrict the privilege or side
effects of a supply chain element with the goal of minimizing the
surface for confidentiality or integrity attacks.

Ephemeral credentials constrain a compromised principal 
by limiting the use of credentials temporally (\textbf{P1}).
Techniques that restrict the interface to an supply chain resource's
environment (\textbf{R2}) or enforce strict environment boundaries (\textbf{R3})
spatially confine a resource so that it may not operate outside its expected scope.

\textbf{{Correctness}} defenses have the goal of ensuring correct \textit{behavior} in
supply chain steps by minimizing the occurrence of errors or
exploitable weaknesses, and some even generating verifiable proofs
of correctness.

All program analysis techniques to enforce expected artifact behavior (\textbf{A3}) examine
an artifact's functionality to detect anomalous functionality, with some techniques like
patch propagation and dynamic/ML-based analysis correcting vulnerable behaviors.
Deterministic steps, on the other hand, enforce expected step behavior (\textbf{S1}) via
specifying prior configurations that define correct and consistent step operations.
Finally, correctness and consistent results across a supply chain DAG (\textbf{T2})
is achieved through reproducibility of an entire DAG, a technique that leverages
defenses for all other supply chain elements.

\insightbox{Each \swsc defense can be characterized by its \textbf{scope}, \textbf{context}, and \textbf{outcome}. We identify six recurring defense patterns, which we map into three high-level security control types: \textit{detective}, \textit{corrective}, and \textit{preventative}.% These patterns enable systematic mapping to shared security objectives.
}
\section{Trends and Future Directions}
\label{sec:future}
Finally, we utilize the insights from above to derive information about the state of \swsc defense practice to identify current trends and future work directions.
These trends can shed light in existing gaps, as well as mis-allocated efforts which can in turn guide new research directions.
In order to identify these gaps, we also utilize supply chain compromise data to ground our study in empiricism.

\subsection{Attack Distribution Across AStRA Elements} To validate the expressiveness of \model model and the practical relevance of our security objectives, We analyze the attacks in the  In-Q-Tel (IQT) lab’s software supply chain compromises dataset\cite{iqt-dataset}, which is a foundational repository used across other \swsc systematization literature such as Ladisa~\ea~\cite{ladisa-sok}.
Though other datasets exist (e.g., Ohm~\ea), this dataset provides crucial information --- the point of compromise --- and various classess of \swsc attacks (e.g., as opposed to malware samples).
The dataset covers 183 incidents with detailed descriptions, categories, references, and timelines.

Building on this, we perform a comparative analysis between \model’s security objectives and the attack classes represented in the dataset. 
Given the natural with the Ladisa~\ea SoK~\cite{ladisa-sok}, we also utilize their \swsc attack taxonomy to create a \textit{(compromise, attack\_vector, security\_objective\_violated)} triple.
Grounding our evaluation in this dataset allows us to examine how documented compromises manifest across \model elements and the overall DAG structure, and to assess whether \model captures the breadth and nuance of adversarial behaviors observed in practice. 
%The complete analysis table is in the paper’s artifacts repository. %We only included attacks with valid references, that has enough published details to understand the compromised supply chains.Due to space reasons, we have made the full table with our analysis available online.\footnote{https://shorturl.at/JnpjG}

%\msm{these should be drawn from the insights and discussions throughout the rest of the paper}

To perform this mapping, we identified the type of compromise each incident and matched it to the affected AStRA element as well as the violated security objective(s). 
For example, if the dataset reported key leakage or unauthorized account access, we labeled it as a principal compromise, if the cause involved a compromised CI/CD service, or build tool, we labeled it as a resource compromise, if it involves running malicious build, we labeled it as a step compromise. 
In cases where the compromise chain spanned multiple elements (e.g., a principal compromise leading to a resource compromise), we recorded all affected AStRA elements involved.
%Based on the affected elements, we assigned the violated security objective(s). 
For example, distributing a malicious version of a package to include remote access to users machines violates the expected behavior of an artifact (A3), and detecting artifacts changes (A1). 
Extra resources security objectives may have helped reducing this risk (e.g. limiting resource privileges and restricting environment interfaces <<R1, R2>> ). 

Our analysis demonstrates that all attacks in the dataset can be represented using the \model elements and mapped to the security objectives defined in our defense-oriented taxonomy. 
Aggregating the incidents by \model  element reveals a pronounced skew in how supply chain compromises manifest. Artifact-level attacks overwhelmingly dominate, accounting for (82\%), reflecting the prevalence of malicious package injection, and dependency hijacking. 
The second common category is resource related compromises (26\%), which indicate adversarial pressure on CI/CD environments, privileged infrastructure, and isolation boundaries.

\subsection{Security Objectives and Attack Trends}
\begin{figure}
    \centering
    \includegraphics[width=\columnwidth, trim=0 20 0 20]{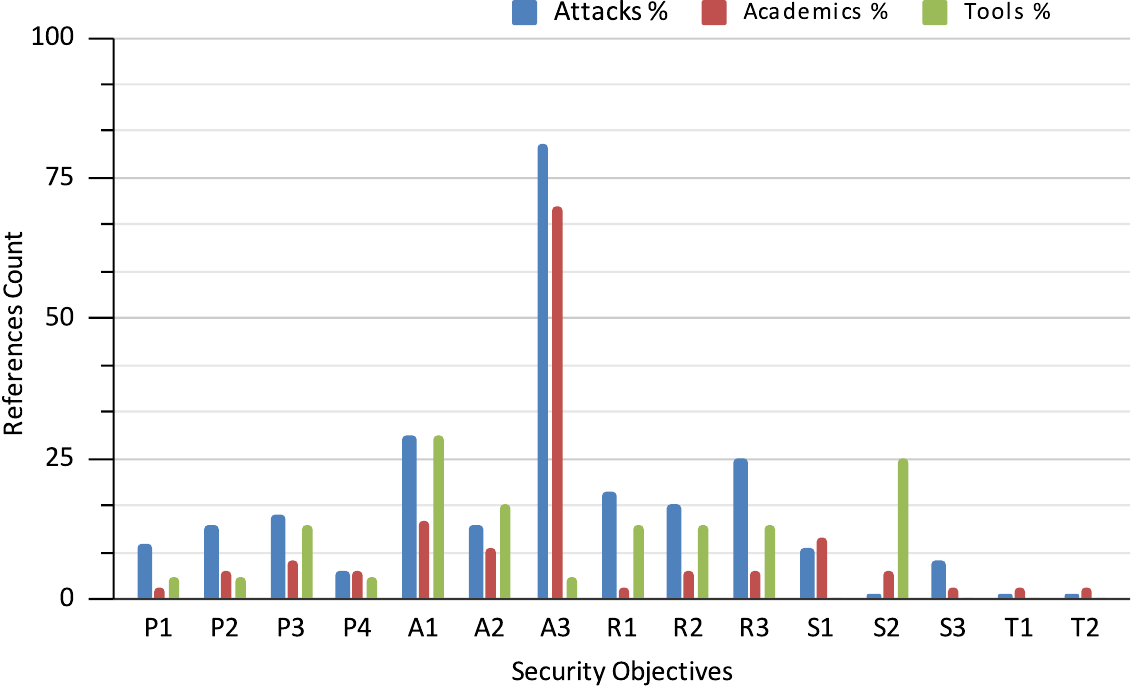}
    \caption{Normalized distribution of \swsc attacks, academic, and tools references  across security objectives
    %\santiago{can you give normalized numbers here?} 
    }
    \label{fig:all_ref_counts}
\end{figure}

By encoding each attack as violations of security objectives, we observe the normalized distribution of attack trends across each security objective, as shown in figure~\ref{fig:all_ref_counts}.
It is clear from the figure that most of the attacks concentrate on violating artifact and resources objectives, while fewer attacks target structural topology or step-level behaviors. 
The most frequently violated objective is A3 (Expected artifact behavior enforced), which appears in 81\% of the incidents, making it by far the most heavily targeted property in real compromises. 
Respectively, Artifact Integrity is the second most targeted property (29\%). Principal objectives (P1–P3) collectively account for 36\% of the incidents, highlighting the persistent role of leaked, reused, or improperly scoped credentials.
This alignment between attack patterns and our security objectives confirms that the objective set is expressive enough to capture known attack behavior and provides the foundation for systematically assessing whether available defenses can reduce the risk of these attacks to succeed.

Although academic defenses and industry tools differ in how heavily they invest in individual objectives, the overall shape of their distributions of defenses in Figure~\ref{fig:all_ref_counts} mirrors the structure of the attack distribution. 
In all three views: attacks, academic work, and tooling: the dominant categories fall under artifacts, followed by resources, then principals, with considerably less activity in steps and topology. 
However, tooling shows notable priority divergences. 
For example, it overinvests in recording step operation (S2), allocating nearly 25\% of its focus to an objective that appears in less than 1\% of attacks. 
This overinvestment likely reflects the high perceived cost of pipeline failures, showing that defense priorities are influenced by the expected operational and economical impact.
A possible explanation is that of the \textit{impact} of a single incident on a particular property --- this asymmetry warrants further research.

 %\insightbox{The fact that attacks, defenses, and tools all show the same ordering of emphasis (artifacts →  resources → principals → steps/topology) indicates that  the way AStRA’s decomposes supply-chain security into objectives is not arbitrary: independent ecosystems of research, practice, and adversarial behavior naturally cluster around the same dimensions.}
\insightbox{Academic defenses align nearly five times closer to real attack patterns than industrial tools, indicating that while research attention better reflects observed risk, more work is needed to bring tooling into alignment with that risk, particularly in enforcing expected artifact and step behavior (A3,S1) and in detecting interference with steps (S3).}

%\santiago{Try to frame this paragarph with the information above. e.g., ``using this empirical data, we now move on to identify gaps in research and defenses and whether they match the trends in practical attacks}
Using this empirical data, we now discuss gaps in the \swsc defense space, and discuss the high-level trend of security objectives addressed in the literature (from Fig.~\ref{fig:all_ref_counts}). 

%In this section, we answer \textbf{RQ4} by outlining future research opportunities in this area.

\subsection{Principal discussion \& Research Gaps}
Table~\ref{tab:sok-table-compact} reveals that no individual approach is sufficient to reduce the possible risks to the principals.
Practical defenses have combined multiple approaches to provide stronger security properties.
A promising combination of One Time Passwords (\eg OTPs ) with multifactor authentication (\eg two-factor authentication~\cite{gh-2fa}) reduces risks in two key areas (\textbf{P1}, \textbf{P3}). 
Despite the highly collaborative nature of today's software development practices, this widely adopted defense combination primarily reduces risks for individual principals, who represent single points of failure. 
Providing further security properties requires layering this approach with others to raise the bar to compromise an entire supply chain via a single principal.

Another promising combination of approaches are short-lived credentials like those provided by Sigstore's Fulcio~\cite{fulcio} and credential transparency: 
 in fact, Fulcio relies on CT~\cite{ct} for the detection of compromise and the added reliability of short-lived credentials issued, reducing risks in three key areas (\textbf{P1-P3}).
To raise usability, Fulcio relies on OpenID Connect~\cite{oidc-connect} and timestamping services (RFC 3161)~\cite{rfc-3161} for credential generation and discovery.
Layering these defenses with additional approaches to address \textbf{P4} would likely not impact adoption costs significantly.

\Paragraph{Research gaps.}
Several approaches in our analysis have potential for addressing Principal risks, although they have been studied only in other application domains.
Threshold schemes, for example, have been applied in other use cases such as private information retrieval (\eg~\cite{wang-pir-splinter}) and peer-to-peer membership control (\eg~\cite{narasimha-p2p-threshold}).
However, to our knowledge, their application in the \swsc has yet to be explored. 
Threshold schemes, for instance, meet several principal security objectives (\textbf{P3, P4}); their applicability to improve the reliability of authorization in the supply chain is a promising future research direction.

In addition, credential transparency is a promising approach to detect credential
compromise. 
However, its limited adoption for \swsc security warrants study to determine its applicability in practice. 
%Similarly, understanding how other defenses are combined with credential transparency remains an open question.

\subsection{Artifacts discussion \& Research Gaps}
Perhaps the most noticeable aspect is the vast prevalence of \textbf{A3} as a security objective within \swsc scholarship, as shown in Figure~\ref{fig:all_ref_counts}.
%Yet, there are other security objectives that remain comparatively understudied.
Work in this domain is predominantly achieved by means of standard program analysis techniques.
Refocusing efforts in other security objectives and techniques in the artifacts domain will likely be a fruitful pursuit.

With regard to techniques, Table~\ref{tab:sok-table-compact} shows that there is no one-stop-shop to achieve all desired objectives, nor are several single approaches enough to achieve individual properties.
Artifact and metadata signing, for example, are both needed to fully achieve \textbf{A1}.
Static and dynamic analysis techniques are often combined for improved results to identify and enforce the expected artifact behavior (\textbf{A3}).

While composing systems to achieve all artifact security objectives may be feasible (e.g., by means of dynamic analysis and metadata signing),  there is no established best practice for doing so, nor have the security implications of such combined systems been studied.
This is a stark contrast with other elements of the model (see, e.g., the resources below), which have garnered significant industry interest in establishing guidance in the supply chain.

\Paragraph{Research gaps.} Some of the approaches evaluated have seen varying degrees of adoption.
This is evidenced by the low adoption of solutions such as artifact and metadata signing in critical areas such as package management~\cite{kuppusamy-nsdi-2016, wwoodruff-gpg, package-signing}.
For hashing, for instance, research on mechanisms such as subresource integrity HTML tags~\cite{Subresource_Integrity} may address known usability issues~\cite{Usable_Checksums}, but more studies are needed on more defenses to understand adoption barriers.

Work establishing better mental models about trust in artifacts like packages through the above properties is lacking.
In particular, security policies for software trust in different ecosystems are currently ad-hoc and do not follow a principled definition based on white or gray literature.
That is, systems such as GPG have provided mechanisms for developers to sign packages, but they rarely develop infrastructure and processes to verify that the right actor signed for the right artifact. 
In addition, there is little research exploring the usability and process challenges on proper definition and enforcement of policies. 

Approaches such as ledger-backed systems face performance and efficiency tradeoffs introduced by consensus algorithms used to help achieve \textbf{A2}. 
Given that these approaches can provide availability and privacy properties~\cite{merrill_speranza_2023} in addition to meeting objectives (\textbf{A1, A2}), and these mechanisms provide strong deterrent to attackers, more research is needed to reduce their deployment cost.
%in a way that breaking its integrity costs millions of dollars even for attackers who can carry out a man in the middle attack

\subsection{Resources discussion \& Research Gaps}
%- Is there a winner? 
As with prior supply chain elements, defense approaches for resources are complementary.
However, unlike other areas of the chain, the long-standing history of computer systems security research has led to better compatibility between the evaluated approaches.

For instance, access control mechanisms are capable of operating at various levels of granularity depending on a resource's operating environment. 
Although implementations such as AppArmor~\cite{apparmor} provide application-level MAC, DAC can provide system-level defenses.
At the same time, filtering mechanisms, such as netfilter~\cite{wool_netfilter_usability} can provide system-level protection in addition to DAC and MAC within the system. 
Virtualization then provides an additional level of isolation in addition to access control and filtering.

We observe that resource defense approaches appear more frequently in industry security best-practice documents that have been authored in response to recent supply chain compromises.
Most notably, the Cloud Native Compute Foundation's Software Supply Chain Best Practices~\cite{cncf-best-practices} and the Secure Software Development Framework (SSDF) by NIST~\cite{nist-ssdf} both describe a set of recommendations for configuring and implementing more secure \swsc resources that rely on defense approaches including MAC, filtering, and virtualization. %using network segmentation and access controls to separate development environments from production environments.%, and to separate components within the build environment, to reduce attack surfaces and attackers’ lateral movement and privilege/access escalation. 
The Supply Chain Levels for Software Artifacts (SLSA)~\cite{slsa}, which provides a framework for incrementally implementing a hardened build pipeline, also includes requirements for resource isolation to ensure that build steps are performed without external influence to meet higher security levels.

\Paragraph{Research gaps.} Academic research analyzing the security of \swsc resources thus far taken
primarily an attack-oriented view (\eg \cite{maloss}). 
Although the design and application of access control and resource isolation techniques have been widely studied in the academic systems community, research dedicated to applying these defense mechanisms in the context of \swsc security is lacking;
we see ample opportunities to build upon defense-oriented academic studies such as Gruhn~\ea~\cite{sec-ci}
and Ohm~\ea~\cite{ohm-sok}.

%%The Secure Software Development Framework (SSDF) by NIST recommends using risk-based approach to adopting secure software development practices.
%Access control and resource isolation techniques mitigate the risk of resources in the supply chain. 
For brevity, we do not extensively evaluate resource defense approaches such as bounds checking (\eg \cite{sfi}), or more recent hardware-based techniques such as trusted execution environments (\eg~\cite{sgx,trustzone,tdx,amd-sev}).
However, we observe in existing best practices guides that such approaches are underexplored in supply chain settings.
In particular, the inclusion of hardware-based attestation primitives along with software/provenance attestation seems like a natural construction to explore.

The focus on reducing risks of building and packaging resources~\cite{sec-ci,ohm-sok,slsa}
leaves an opportunity for future research to explore existing and novel approaches for other \swsc use cases.
More research is needed on the adoption of industry best practices.
%this can reduce the risk of tampering with resources and steps in the \swsc. 

% 
%SLSA level 3 and 4 requires the build environment to be isolated in such a way that the build cant access any secrets, the build output must be identical whether or not the cache is used, and different builds that overlaps or are subsequent can't influence each others. To help achieve these properties, SLSA requires the build environment to be  ephemeral like containers or VMs, so it can only be used for this build, and not reused from prior build or not used again for future builds.
%In earlier versions of SLSA, builds must be hermetic to achieve level 4; this requires build steps to run with no network access, which means that all dependencies should be fetched before the build steps occurs. To apply this requirement, SLSA considered using containers to prevent internet access sufficient. However, this requirement is considered hard to achieve in practice and labeled as best effort, given that the builds should be hosted in infrastructure like GitHub Actions, Google Cloud Build. Having a good rational reasoning behind such strict requirement, could have led to it being used more in practice. In the newest SLSA version,1.0 , this requirement has been omitted and labeled as future work which may or may not be included in L4. 

\subsection{Steps Discussion \& Research Gaps}
The majority of step defenses work has focused on mechanisms for recording step metadata to detect tampering.
Xia~\ea~\cite{SBOM-study}, for instance, conducted a study that focuses on current SBOM practices, tooling and concerns from SBOM practitioners’ perspectives. 
The study shows lack of SBOM adoption and problems in enabling SBOMs benefits, due to various concerns 
regarding SBOM generation, distribution, consumption processes. 
% Concerns on SBOM generations includes: SBOM standards require further consensus, standardization, and extension points, SBOM generation is belated and not dynamic through the software development stages,  lack of SBOM data validation and integrity, lack of maturity in SBOM tooling, more reliable, user-friendly, standard-conformable, and enterprise-level SBOM tools is needed. 

% Concerns on SBOM distribution includes barriers to SBOM distribution introduced by proprietary and sensitive information in SBOMs. 
% Concerns on SBOM consumption includes lack of systematic consumption-scenario driven design of SBOM features, and lack of  market awareness and promotion and good value propositions for SBOM adoption.
%Given these concerns, the study suggests three prerequisites to increase SBOM adoption and SBOM enabled benefits: higher quality SBOM generation, clearer benefits and use cases in SBOM consumption, and lower barriers in SBOM sharing. 
%
%{\textbf{Formal verification}} mechanisms leverage mathematical principles, such as discrete logic,
%to check whether an algorithm or design meet specific properties or behavior. 
%Typically, this approach
%may be used to reduce risks to steps by checking the behavioral correctness of the resources used
%to carry out a step (\textbf{S1}).

\Paragraph{Research gaps.}
In both academic and gray literature, we observe a general lack of studies on preventing or reducing risks that affect the correct execution or operation of \swsc steps.
Formal verification, for instance, has been widely explored by the programming language and
software engineering communities, though its application in the context of
the software supply chain remains underexplored.
%\Paragraph{Verifiable Systems to Ensure Correctness}
Furthermore, the adoption of formal verification in resources for steps is  
a promising research opportunity.

Although the design of consensus-based step defense mechanisms has been the focus of
an increasing number of academic studies, their practical applicability and adoption
costs remain relatively underexplored. 
Similarly, the design, development, and adoption of step containment approaches at large is an area that warrants further study.

\subsection{Other Salient Points \& Research Gaps}
The \swsc DAG shares many characteristics with multi-party, distributed systems.
Given the focus on securing individual supply chain elements, \swsc security would greatly benefit
from more applying knowledge and from distributed systems, such as fault tolerance and replication, to address transitive risks.

%\msm{incorporate some of the commented text about systems reliability etc.}

\comment{further from systems research on redundancy, we observe that while some mechanisms for
replication are applicable in the \swsc, the stronger property of \textit{fault tolerance}, in which
a system may tolerate or even recover from unexpected changes such as modifications to the supply chain DAG
structure~\cite{fault-tolerant-os, fault-tolerant-sys} remain unexplored. One possible means to achieving 
this property may be by substituting an individual compromised or faulty \swsc element
with an equivalent element before re-attempting to execute a given step, though further research and engineering
are necessary to better understand the impact of such fault recovery mechanisms to the security, usability,
and performance of a \swsc.

No single point of failure: Reduces the risks of failure in a \swsc due to a missing step or resource.
This is achieved through \textit{redundancy} mechanisms~\cite{fault-tolerant-os,fault-tolerant-sys}.

{\textbf{Resource replication}} addresses risks in a step caused by a single point of failure such as a removed or 
problematic resource that the particular step in the \swsc graph depends on. For instance, in the Debian or Ubuntu package ecosystem, principals
may configure their dependency fetching steps to use ``mirrors''~\cite{apt-mirror}, which replicate and regularly synchronize with a central package
registry to provide a fallback mechanism in case that a chosen package registry becomes unavailable or otherwise inaccessible (\textbf{T2}). In the case of left-pad incident ~\cite{}, this technique could have helped the impacted builds to succeed.

Attack propagation prevented: Limits the risks to a particular step, resource or principal by blocking or avoiding
an interaction or transitive relationship between two vertices in a \swsc graph.

{\textbf{Dependency reduction}} is a defense technique that intentionally modifies a supply chain DAG itself in order to
block or avoid transitive relationships between two vertices, such as principals, resources and steps. One solution specifically developed
to address compromises in artifacts that are built using a compiler (\ie a resource) such as GCC is Bootstrappable builds~\cite{bootstrappable-builds,bootstrappable_builds_website}.
The key idea behind Bootstrappable builds is that the smaller the so-called ``seed'' compiler is for a project, the lower the risks of
compromise are for the project since a seed can be individually inspected, increasing the transparency of the dependencies.
Reducing a resource dependency to a seed ultimately limits the transitive relationships between a step and a resource (\textbf{T3}).

Topology non-equivocation: All principals have the same view of the topology at a given point in time, this includes the set and order of steps in the topology.
Further, mechanisms to provide a transparent, tight binding between policies and their evaluation will allow for stronger insights on the security posture of an artifact in context (\textbf{T4}).
}

\Paragraph{Reducing Barriers to Adoption.} Originally, we intended to evaluate the 
tradeoffs and costs to adopt mechanisms for \swsc security.
Unfortunately, we found there are very little literature and formal studies on how to deploy a particular approach. 
These studies should cover human factors (e.g., training), usability, and system requirements (e.g., extra servers).
Thus, we envision a green field for research in \textit{Usable \swsc security}.

\Paragraph{Moving beyond modeling threats.} 
%Prior research covers \swsc threats extensively. 
%As mentioned earlier, there is plenty of work on modeling, studying and taxonomizing risks, yet how these risks map to 
%mitigations and defenses is generally underexplored.
Our objective-driven model is a first step towards systematically mapping high-level security objectives and patterns to defenses.
We see an opportunity to expand upon our work to design novel adoption-ready \swsc defenses.
Beyond our preliminary work, we envision future research that incorporates holistic 
cost-benefit analysis frameworks for supply chain risk mitigation with the goal of increasing adoption of 
supply chain defenses.

\Paragraph{Understanding the impact \& security of AI-generated software.} Generative AI
is playing an increasingly central role in software development. According to a recent survey~\cite{github-survey-2023},
about 92\% of developers in the U.S. rely on AI-based tools. This rapid shift presents ample future research 
opportunities to study the security implications of AI-generated code and address software and supply chain security in this space.

\insightbox{
There are four significant opportunities:
  \textbf{1)} Studying defense adoption and usability.
\textbf{2)} Developing deployable defenses grounded in human factors.
\textbf{3)} Exploring foundational tools (e.g., %access control, isolation, formal verification, bounds checking, and 
trusted execution environments) in supply chain contexts.
\textbf{4)} Formalizing trust models and policies for artifact provenance.
}
%Conclusions:
%\begin{itemize}
%    \item If we can reduce risks in principals, steps, resources, topo, we can greatly reduce risks to artifacts.
%    \item Improving education, automation, and regulation?
%    \item 08/04/2023: we wanted to do UX research, and we couldn't (be snarky in ways) + Hash literature we found, surprising information about human factors that lead developers to not use a mechanism. Quotes and interviews regarding public CI systems, why need ACL for some actions and operations? Tie back to one of the problems: people don't know how to reason/understand the risks on the SWSC. (Attacks: reactive, risks: proactive)
%
%\end{itemize}

\section{Conclusions}

This paper reframes the security of the software supply chain from an attacker's checklist to a defender’s blueprint. 
By modeling chains as DAGs of principals, artifacts, resources, and steps, we distill fifteen cross-cutting security objectives and map academic and industrial mechanisms onto them, exposing three recurring defense patterns.
We further explore how these patterns surface possible secure configurations, and note trends in existing literature in the area of \swsc.
With this in mind, we also surface promising but underexplored lines of work.

By making shortcomings explicit and placing existing defenses on a common scaffold, we provide researchers with a roadmap of open problems, practitioners with guidance for defense prioritization, and policymakers with levers for measurable requirements. 
We hope that this objective-driven perspective catalyzes a new generation of holistic, composable defenses capable of keeping pace with the dynamic, federated nature of today’s software supply chains.

%This document reframes the security of the software supply chain from an attacker's checklist to a defender’s blueprint. 
%By modeling chains as DAGs of principals, artifacts, resources, and steps, we distill fifteen cross-cutting security objectives and map academic and industrial mechanisms onto them, exposing three recurring defense patterns, limitation, correctness, and transparency, and clarifying the contexts in which each succeeds or fails. 
%
%Our analysis shows that credential hardening and resource sandboxing are mature and widely deployed, yet cover only a subset of objectives. 
%Similarly, we find that few defenses reason about the transitive structure of the DAG, provide reproducible end-to-end provenance, or confront human-centric adoption barriers. 
%
%Finally, we surface promising but underexplored lines of work like threshold authorization and credential transparency for principals, formal verification and deterministic execution for steps, layout-integrity checks for software supply chain DAG structure, and usable composition of artifact-level techniques. 
%
%By making shortcomings explicit and placing existing defenses on a common scaffold, we provide researchers with a roadmap of open problems, practitioners with guidance for defense prioritization, and policymakers with levers for measurable requirements. We hope that this objective-driven perspective catalyzes a new generation of holistic, composable defenses capable of keeping pace with the dynamic, federated nature of today’s software supply chains.
%
%\bibliographystyle{IEEEtran}
\section*{Ethics Considerations and Open Science}

%This research primarily focused on capturing information from previous literature and taxonomizing it.
%As such, no \emph{populations or stakeholders} were directly affected by this work.
%Explicitly, we collected 

To support open science principles and ensure transparency, artifacts produced in this study, including datasets and data collection scripts, are released under an open source license and made publicly accessible. This enables verification and reproducibility of the results. Ethical considerations were taken into account throughout the work: no proprietary or sensitive data were used, and all third-party components were used according to their respective license terms.

\section*{Acknowledgments}
This material is based on work supported in part by the Air Force Office of Scientific Research (AFOSR) under award number FA9550-24-1-0015; by the National Center for Transportation Cybersecurity and Resiliency (TraCR) under grant numbers 69A3552344812 and 69A3552348317; and by a grant from the Digital Infrastructure Insights Fund, a fiscally sponsored project of Aspiration.

\bibliographystyle{abbrv}

\begingroup
\setlength{\bibsep}{0pt} % remove spacing between items
\footnotesize
\bibliography{references}
\endgroup

%{\footnotesize
%\bibliography{references}}

%\input{case-studies}
\appendices

%\subsection{Generality and Fidelity of \model}

\section{Generality and Fidelity of \model}
\label{sec:evaluation}

To demonstrate that AStRA is not merely a theoretical exercise but a necessary framework for defense, we validate its Generality (its ability to model diverse ecosystems) and its Fidelity (its ability to explain real-world attacks).

\subsection{Generality}
%1st how you gonna evaluate it?
We evaluate the generality of our model using two case studies widely studied in both scientific and gray literature: Left-pad (npm)~\cite{leftpad-npm} and SolarWinds~\cite{solarwinds-fireeye}.
%We chose the supply chains of both SolarWinds~\cite{trebuchet} and left-pad~\cite{left-pad} as they will become useful when evaluating completeness.

\begin{figure*}[h]
        \centering
        \includegraphics[width=\linewidth]{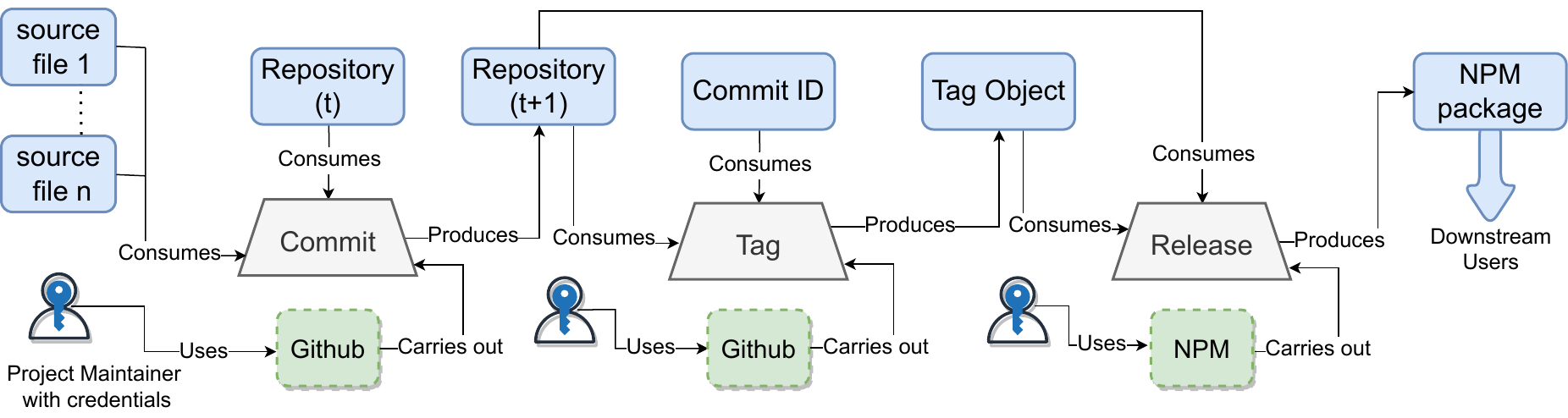}
    \caption{\footnotesize The supply chain DAG of left-pad, the commit step can occur multiple times before tagging the repo, and all the steps in figure occur for each release.}
    \label{fig:left-pad}
\end{figure*}

\noindent\textbf{Left-pad} is a single-developer npm workflow, we determined the supply chain for it by analyzing its source code repository~\cite{left-pad-repo}.
Succinctly put, the left-pad is a linear source code repository that, when the main developer deems appropriate, uses git tags to mark the release of a version. No automated pipeline is used to release, suggesting that the developer uses a script (i.e.,~\cite{npm_publish} to upload the corresponding version to NPM. The same principal is in charge of pushing the resulting binary to a registry for consumers to download. Figure~\ref{fig:left-pad} represents the AStRA topology for Left-Pad.

\begin{figure*}[h]
        \centering
        \includegraphics[width=\linewidth]{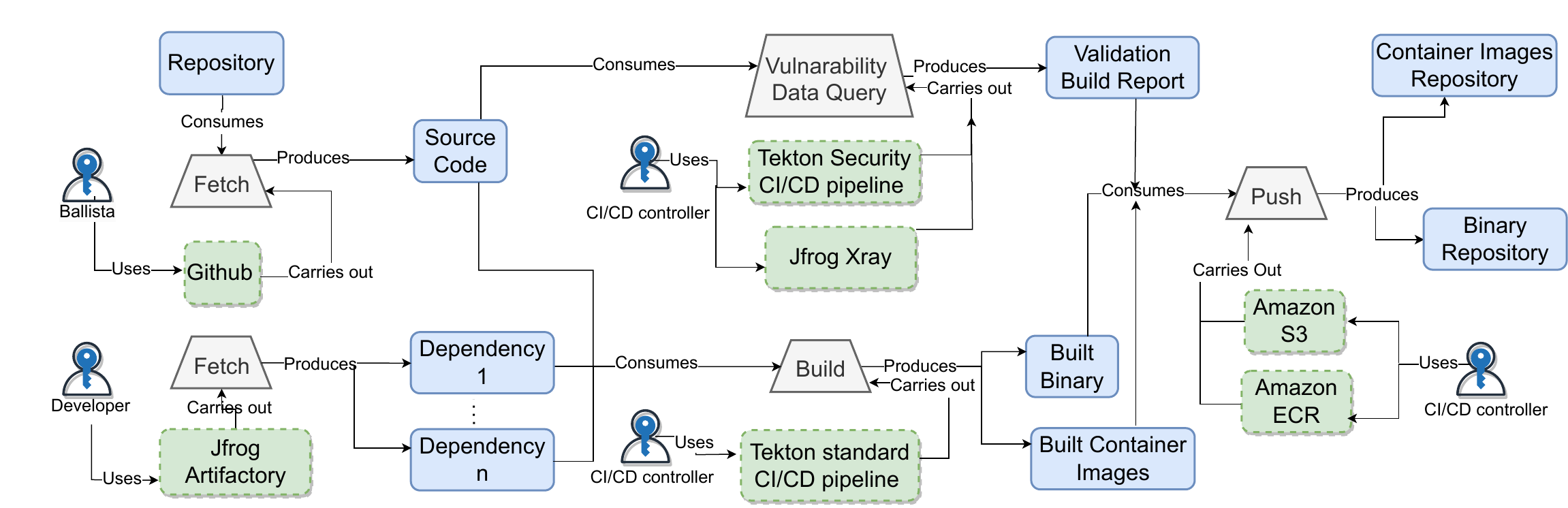}
    \caption{\footnotesize The supply chain DAG of SolarWinds build system.}
    \label{fig:solarwinds-trebuchet}
\end{figure*}

\noindent\textbf{SolarWinds} is a multi-principal Tekton-based CI/CD pipeline that uses a combination of hosted services (e.g., GitHub) as well as self-hosted tools and infrastructure (e.g., Tekton) to provide a container build pipeline. Of particular interest is the use of different principals to report dependencies as a separate step. This is done this way because the automated pipeline pre-fetches the sources and installs them on the build infrastructure at the time of build. One possible reason for this is that developers need clearance and oversight when selecting third-party components.
Separately, developers build their own source code and host it in a social coding platform, which is then tagged and released for build by the Tekton pipeline manager.
The same principal is in charge of pushing the resulting binary to a registry for consumers to download. Figure~\ref{fig:solarwinds-trebuchet}
represents the AStRA topology for SolarWinds. 

\begin{figure*}[h]
        \centering
        \includegraphics[width=\linewidth]{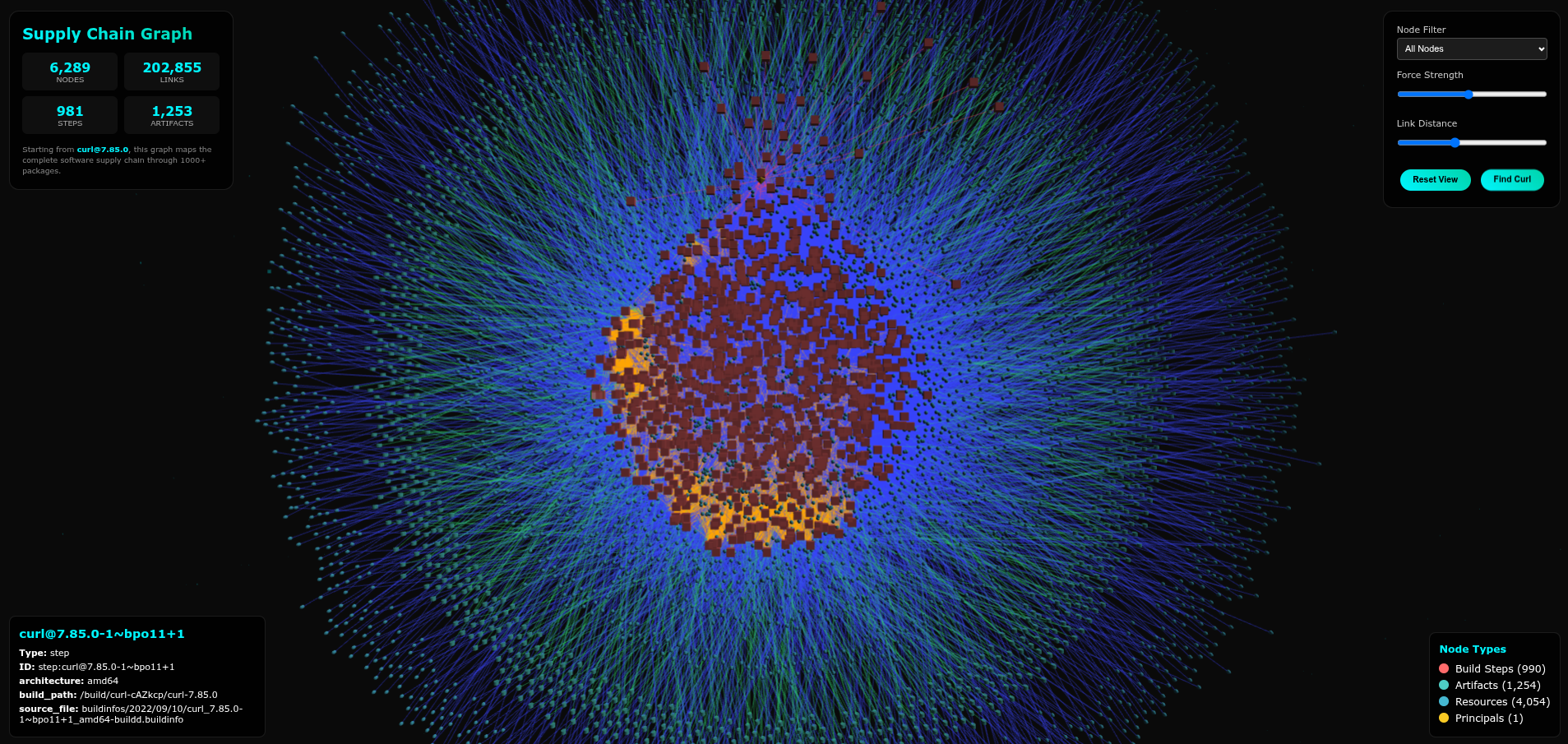}
    \caption{\footnotesize An automated supply chain representation for 990 debian builds.}
    \label{fig:debian-astra}
\end{figure*}

\noindent\textbf{Debian’s Reproducible Builds}
Large software ecosystems naturally produce sizable AStRA graphs, motivating dedicated tooling for their construction and analysis. To evaluate \model model practicality at scale, we have implemented a prototype that automatically generates AStRA graphs from Debian’s reproducible-builds metadata~\cite{reproduciblebuilds_buildinfo}, which covers more than a decade of Debian packaging activities. The resulting graph contains approximately 9 million nodes ($\approx$ 7 M artifacts, 285 K resources, 1.3 M steps, and 1.2 K principals) and 407 M edges, with an average of $\approx$ 302 edges per step. These aggregate statistics confirm that AStRA can realistically represent production-scale supply chains and that such graphs are amenable to automated processing using existing graph-analysis frameworks (e.g., NetworkX, Neo4j).
To show generality, we present an automated AStRA representation of a Debian buildserver using buildinfo~\cite{reproduciblebuilds_buildinfo} metadata on Figure~\ref{fig:debian-astra}. 

%is a software company that offers an IT performance management and monitoring system called Orion. 
 %we made these assumptions about ... every single actor that is not well documneted is considered a seperate actor because we assume that it is a stronger configuration
%3rd analysis parag
% For SolarWinds \swsc arrangment we end up with with a topology that has this ammount of nodes of this different types . Table .. represent general graph structure properties for the model for SolarWinds case.

%\paragraph{Left-Pad}is popoular npm JavaScript package (17 lines of code) that pre-pends zeroes or spaces to strings. Despite its size, the package had an average  of weekly downloads that exceeds one million~\cite{leftpad_npm}. Using the metadata available at the package repository on github~\cite{leftpad_github}, we were able to model the left-pad \swsc using our conceptual model. To do so, we used the commit history of the package, and for each step mentioned in the history, we determined the four elements of a \swsc that is included in the step and their relations. After getting the steps of the software supply chain, we chained the steps in a chronological order to get the software supply chain \textit{topology}. Table \ref{tab:left-pad} represents the left-pad \swsc topology graph.
Modeling SolarWinds, Left-pad, and Debian builds supply chains using the same four-element abstraction (Principals, Steps, Resources, and Artifacts) and without altering the schema, demonstrates that AStRA generalizes across manual and automated, single- and multi-actor workflows.

\subsection{Fidelity}
%\subsubsection{Fidelity}
%1st how you gonna evaluate it?+ %2nd what you chose
To evaluate this property, and given that we have proved the generality of our model, we will use it to analyze widely-known supply chain attacks on both SolarWinds and left-pad supply chains. 
We show how each compromise materializes within an AStRA model.

\noindent\textbf{Left-Pad} was un-published by its author, as part of a dispute over a package name with the administrators of the npm package manager~\cite{leftpad}. Since this was done without warning to the developers of the projects that depended on it, this affected many thousands of projects and npm observed hundreds of build failures per minute for the dependent projects and their dependents~\cite{leftpad}. 
Based on our conceptual model, the artifact removal imposes a topological risk on all the software supply chain topologies that used left-pad as a direct or transitive dependency. This topology risk can be reduced using resource replication techniques to prevent a dependency from being a single point of failure in the build steps. %(\textbf{T2}).

%Developers of software supply chain defenses can benefit from analyzing their \swsc topology to ensure  their defenses are meeting certain security objectives. 
Open source package managers can benefit from analyzing the \swsc topologies of the software they publish, to prevent incidents such as Left-Pad. Packages with high out-degrees can represent single point of failure, while packages with long dependency chains increase transitive risk.

\noindent\textbf{SolarWinds} In 2019, a sophisticated adversary used the SolarWinds Orion platform to plant stealthy backdoors in the networks of thousands of US government agencies and companies around the world~\cite{reuters_solarwinds, solarwinds-report}. Although ``SolarWinds'' is used to refer to a single compromise, it was a campaign with multiple stages. 
The attackers injected malicious code known as Sunburst (aka Solorigate) into Orion during the build process. 
The report describes the reuse of credentials on the principals who carry out the build process. In this case, the nodes affected in Figure~\ref{fig:solarwinds-trebuchet} were the CI/CD controller principals.

\comment{Our evaluations (\S\ref{sec:principals}-\S\ref{sec:topology}), summarized in Table~\ref{tab:AStRA},
provide key insights into a variety of defenses that address specific threats to individual \swsc elements.
While our conceptual model may seem to oversimplify the \swsc at first glance, we
argue that the abstractions the model provides aid in identifying common \swsc
defense patterns.
}

Further, the properties of the network structure can improve the analysis of assets and elements involved in a supply chain. For example, identifying principals with high out-degree would suggest an excess of privilege (and thus a higher compromise surface).
Similar analysis on out-degree can be used to indicate artifact, resource, or even step criticality.
%Given the released information, and although we do not know the authorization mechanism used, methods like Multi-factor authentication could have made it harder for SolarWinds attackers to authenticate themselves at the initial entry into the build system(P3). Also, techniques like Ephemeral credentials could have restricted the compromise impact and prevented the attackers from staying too long in the system (P1).  Techniques like threshold authorization could have helped narrowing the impact of the compromise and may prevent the threat principals from performing building steps (P3,P4). 

In both attacks, \model correctly localizes the source of the compromise (credential misuse or single-point dependency) and can trace its transitive effects across the supply chain.

This validation is critical for the systematization of this paper, as this allows us to confirm that Principals, Artifacts, Resources, Steps, and the DAG Structure create a model for defense. Consequently, the taxonomy we present is explicitly structured around these five elements. By securing these five objectives, we can mathematically assert coverage over the known classes of supply chain attacks.

\section{SoK Methodology (Complements~\S\ref{secsec:sok-structure})}
\label{app:methodology}
From Phase 1 of our systematization, we identified 671 potential software supply chain security papers. Table~\ref{tab:conferences} summarizes the venues represented in this corpus and lists each conference along with the number of papers contributed by that venue.
\begin{table}[h!]
  \vspace{-6pt} % reduce space above table
  \centering
  {\scriptsize
  \caption{Conferences used to seed our search.}
  \vspace{-3pt} % reduce space between caption and table
  }
  {\scriptsize
  \setlength{\tabcolsep}{4pt} % reduce horizontal padding in cells
  \begin{tabularx}{\columnwidth}{@{}>{\raggedright\arraybackslash}X c@{}}
    \toprule
    \textbf{Conference} & \textbf{Paper Count} \\ \midrule
    IEEE Symposium on Security and Privacy (US \& Europe)           & 97  \\
    USENIX Security Symposium                                       & 137 \\
    ACM Conference on Computer and Communications Security (US \& Asia) & 138 \\
    Network and Distributed System Security Symposium               & 10  \\
    Annual Computer Security Applications Conference                & 33  \\
    IEEE Computer Security Foundations Symposium                    & 8   \\
    European Symposium on Research in Computer Security             & 35  \\
    Privacy Enhancing Technologies Symposium                        & 16  \\
    International Symposium on Research in Attacks, Intrusions and Defenses & 16  \\
    Detection of Intrusions and Malware, and Vulnerability Assessment & 6   \\
    International Conference on Software Engineering                & 109 \\
    Software Engineering in Practice (ICSE track)                   & 12  \\
    Mining Software Repositories                                     & 26  \\
    USENIX Annual Technical Conference                               & 8   \\
    Networked Systems Design and Implementation                      & 12  \\
    Operating Systems Design and Implementation                      & 8   \\ \midrule
    \textbf{Total} & \textbf{671} \\ \bottomrule
  \end{tabularx}
  }
  \label{tab:conferences}
\end{table}

\begin{figure}[htbp]
    \centering
    \includegraphics[trim=0 35 0 30, width=.95\linewidth]{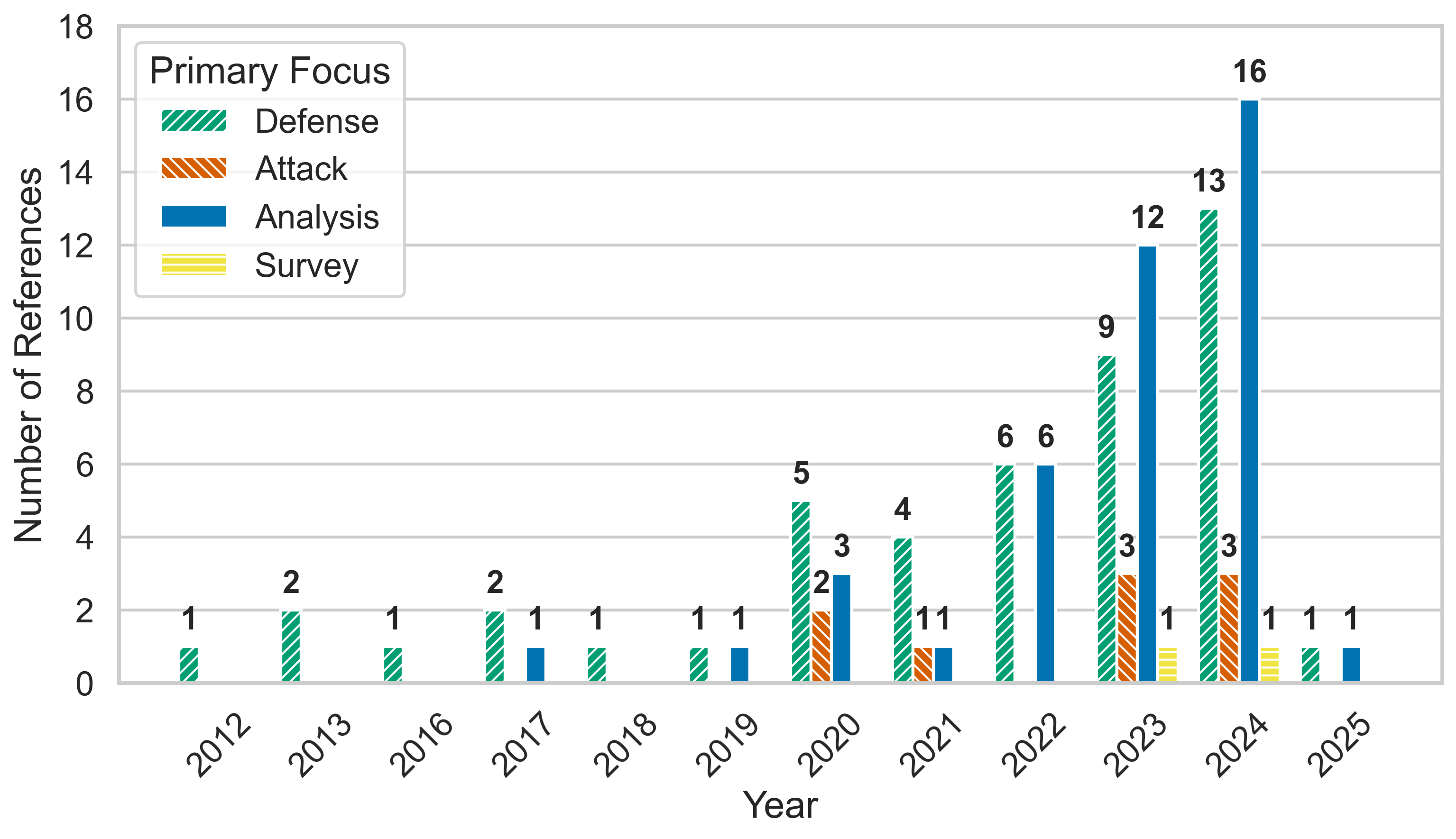}
    \caption{Reference counts by year and primary focus category (Defense, Attack, Security Analysis, Survey) in the software supply chain literature.}
    \label{fig:ref_counts_year}
\end{figure}

Fig.~\ref{fig:ref_counts_year} shows the yearly distribution of references categorized by their primary focus, highlighting trends in academic attention within the software supply chain security field. We can see an overall growth in this area, with initial papers focusing on defenses, while most of the papers in recent years tend to do analysis (e.g., measurement studies of the state of security in the supply chain ecosystem). This is representative of a mature ecosystem, where researchers can measure the adoption of security practices. Consistent with this maturity, we can see that surveys and SoKs started in 2023. As mentioned above, a previous SoK focused on attacks~\cite{ladisa-sok}, and a more recent one~\cite{sok-docker} focuses only on docker attacks and defenses. Our goal is to complement these previous efforts by focusing on defenses against a wide range of supply chain threats. 

Table~\ref{tab:appendix_extra_works} contains 6 well-known papers from industry literature, and 24 tools/frameworks that were added to our corpus of academic literature.
\begin{table}[ht]
\centering
\caption{Defense works extracted from gray literature and industry tools.}
\begin{tabular}{|p{3.5cm}|p{2cm}|p{2cm}|}
\hline
\textbf{Name} & \textbf{Reference} & \textbf{Type} \\ \hline

\textbf{SCITT (IETF)} & \cite{scitt} & White Paper \\ \hline
\textbf{NIST SP 800-204D} & \cite{nist-800-204D} & White Paper \\ \hline
\textbf{CIS Software Supply Chain Security Guide} & \cite{cis-sscs-guide} & White Paper \\ \hline
\textbf{Linux Handbook: The Software Release Life Cycle in DevOps:
Why Reinvent The Model?} & \cite{srlc-linux-handbook} & White Paper \\ \hline
\textbf{CNCF: Software Supply Chain Best Practices} & \cite{cncf-best-practices} & White Paper \\ \hline
\textbf{PEP 480 – End-to-end signing of packages} & \cite{PEP480} & White Paper \\ \hline

\textbf{Fulcio} & \cite{fulcio} & Tool \\ \hline
\textbf{Certificate Transparency} & \cite{ct} & Tool \\ \hline
\textbf{GitHub repository roles} & \cite{gh-repo-roles} & Tool \\ \hline
\textbf{NPM roles} & \cite{npm-roles} & Tool \\ \hline
\textbf{Yubico (hardware MFA)} & \cite{yubico} & Tool \\ \hline
\textbf{Git commit signing (GPG)} & \cite{git-gpg} & Tool \\ \hline
\textbf{GitHub commit signing} & \cite{github-commit-signing} & Tool \\ \hline
\textbf{Maven Maven GPG plugin} & \cite{apache-maven-gpg} & Tool \\ \hline
\textbf{Docker Content Trust} & \cite{docker-image-signing} & Tool \\ \hline
\textbf{Binary Transparency on Firefox} & \cite{ff-ct1} & Tool \\ \hline
\textbf{Sigstore} & \cite{sigstore} & Tool \\ \hline
\textbf{Scorecards} & \cite{scorecards} & Tool \\ \hline
\textbf{Coverity static analysis} & \cite{coverity} & Tool \\ \hline
\textbf{AppArmor} & \cite{apparmor} & Tool \\ \hline
\textbf{SELinux} & \cite{selinux} & Tool \\ \hline
\textbf{eBPF filtering} & \cite{ebpf} & Tool \\ \hline
\textbf{AMD64 Hardware virtualization} & \cite{amd-svm} & Tool \\ \hline
\textbf{Intel Virtualization Technology} & \cite{intel-vtx} & Tool \\ \hline
\textbf{in-toto} & \cite{in-toto} & Tool \\ \hline
\textbf{Trivy} & \cite{trivy} & Tool \\ \hline
\textbf{Supply-chain Levels for Software Artifacts (SLSA)} & \cite{slsa} & Framework \\ \hline
\textbf{Software Bill of Materials (SBOM)} & \cite{sbom} & Framework \\ \hline
\textbf{System Package Data Exchange (SPDX)} & \cite{spdx} & Framework \\ \hline
\textbf{System Package Data Exchange (SPDX)} & \cite{spdx} & Framework \\ \hline
\end{tabular}
\label{tab:appendix_extra_works}
\end{table}

\section{Taxonomy (Complements~\S\ref{secsec:systematization})}
\label{app:taxonomy}

\begin{table*}[]
\caption{Supply chain defense techniques mapped to security objectives and defense patterns. \fullcirc~denotes a technique that meets an objective and falls under a defense pattern. \halfcirc~denotes conditional/limited fulfillment.  \text{\warning} denotes a potential conflict with other objectives. \checkmark denotes a defense technique that was developed specifically for the \swsc. }
\footnotesize
\centering
%\begin{tabular}{|p{3.4cm}|P{1.8cm}|P{1.4cm}|P{1cm}|P{0.3cm}P{0.3cm}P{0.3cm}P{0.3cm}|P{0.2cm}P{0.2cm}|P{0.2cm}P{0.2cm}|P{0.2cm}P{0.2cm}|}
\begin{tabular}{|p{3.4cm}|P{1cm}|P{1.8cm}|P{1.4cm}|P{0.3cm}P{0.3cm}P{0.3cm}P{0.3cm}|lr|lr|lr|}
\hline
\multirow{3}{*}{\textbf{{AStRA Element}}} & \multirow{3}{*}{\makecell{\textbf{Supply} \\ \textbf{Chain} \\ \textbf{Specific?}}} & \multirow{3}{*}{\makecell{\textbf{Academic} \\ \textbf{References}}} & \multirow{3}{*}{\makecell{\textbf{Industry} \\ \textbf{References}}} & \multicolumn{4}{c|}{\multirow{3}{*}{\textbf{{Security Objective}}}} & \multicolumn{6}{c|}{\textbf{{Defense Pattern}}} \\
    &                   &                &   &     &     &     &     &     \multicolumn{2}{P{1.1cm}}{Detective} & \multicolumn{2}{P{1.1cm}}{Corrective} & \multicolumn{2}{P{1.3cm}|}{Preventative} \\
&                   &                 &  & & &  & & \rotatebox[origin=c]{90}{Transparency} & \rotatebox[origin=c]{90}{Integrity} & \rotatebox[origin=c]{90}{Graceful degr.} & \rotatebox[origin=c]{90}{Redundancy} & \rotatebox[origin=c]{90}{Limitation} & \rotatebox[origin=c]{90}{Correctness} \\
\hline
 \textbf{Principals} &                   &                 &  & \textbf{P1} & \textbf{P2} & \textbf{P3} & \textbf{P4} & 
& & & & & \\
ephemeral credentials        &     &   \cite{Newman2022}          &      \cite{fulcio}             &   \fullcirc     &         &         &         &         &           &             &         &      \fullcirc                           &                    \\
secret scanning   &  &  \cite{zhou_hey_2025}                 &                   &      &    \fullcirc     &         &         &    \fullcirc     &                               &                               &                               &                               &                               \\
credential transparency    &   &   \cite{Newman2022}                &     \cite{ct}              &       &     \fullcirc    &         &         &    \fullcirc     &         \fullcirc                      &                               &                               &                      &             \\
role separation            &   &    \cite{Github-CI-Security}               &         \cite{gh-repo-roles, npm-roles}       &         &         &   \fullcirc      &      &         &                               &                  \fullcirc             &                               &                               &                               \\
multi-factor authentication   &      &             &   \cite{yubico}     &        \halfcirc      &         &  \fullcirc       &         &         &             \fullcirc                  &                               &                               &             \halfcirc                  &                               \\
separation of privilege &      &         &        \cite{cncf-best-practices}     &         &         &         &    \fullcirc     &         &                               &                               &                  \fullcirc             &                               &                               \\
threshold authorization     &    &    \cite{TorresArias2019, Newman2022}         &          &         &         &     \fullcirc    &    \fullcirc     &        &                               &              \fullcirc                 &           \fullcirc                &                            &                        
\\[0.2em]
\hline
\textbf{\rule{0pt}{1.2em}Artifacts}           &      &             &                   & \textbf{A1} & \textbf{A2} & \textbf{A3} &     &     &                               &                               &                               &                               &                               \\
hashing       &     & \cite{vu_towards_2020} &              & \halfcirc       &         &       &         &         &        \fullcirc                         &         &                              &                                &                               \\
artifact signing    &     &     \cite{Newman2022}       &  \cite{git-gpg, github-commit-signing, apache-maven-gpg}   & \halfcirc            &       &        &         &    \fullcirc     &         &                               &                               &                               &                                                         \\
metadata signing     & \checkmark        & \cite{karthik_uptane_nodate}            &        \cite{docker-image-signing}         & \fullcirc       & \halfcirc       &        &        &   \halfcirc      &      \fullcirc                         &                               &                               &                             &                               \\
artifact transparency   & \checkmark     & \cite{Newman2022, merrill_speranza_2023} &       \cite{ff-ct1, sigstore, scorecards} & \fullcirc       & \halfcirc       &         &         &    \fullcirc     &         \fullcirc                      &                               &                               &                               &                             \\
ledger-based systems    &      &   \cite{contour_2018}      &              &    \fullcirc     &  \fullcirc        &       &         &    \fullcirc     &         \fullcirc                      &                               &                               &                             &                             \\
static analysis   &   & \cite{gobbi_poster_2023, noauthor_argus_nodate, huang_vmud_2024, huang_donapi_nodate, woo_v1scan_2023, noauthor_product_nodate, ladisa_feasibility_2023, Sofaer_RogueOne, Jia2024, ladisa-java, jarhead} &   \cite{coverity}   &         &         & \halfcirc       &         &         &                               &                               &                             &                               &         \halfcirc                      \\
patch propagation    &  \checkmark &   \cite{dunlap_vfcfinder_2024, Luo2024, spider}      &              &    \text{\warning}  &        &    \fullcirc     &         &         &                               &                               &                             &                           &              \fullcirc                 \\
dynamic analysis   &   & \cite{bytes_fieldfuzz_2023, noauthor_poster_nodate, HARP, he_rapidpatch_nodate, pantelaios_fv8_nodate, wen_5g-spector_2024, ntousakis_demo_2021, huang_donapi_nodate, ladisa_feasibility_2023, Sofaer_RogueOne} &               &       \text{\warning} &         & \halfcirc       &         &         &                               &                               &                              &              &            \fullcirc                   \\
ML-based analysis    &      &   \cite{ladisa_feasibility_2023, nguyen_up2dep_2020, wyss_wolf_2022, Anomalicious, Jiang_BinAI, Jia2024, jarhead}     &              &    \text{\warning}  &        &    \fullcirc    &         &         &                               &                               &                               &                           &                          \fullcirc      
\\[0.2em] 
\hline
\textbf{\rule{0pt}{1.2em}Resources}           &                   &      &          & \textbf{R1} & \textbf{R2} & \textbf{R3} &     &     &                               &                               &                               &                               &                               \\
mandatory access control   &       &           &     \cite{cncf-best-practices, selinux, apparmor}           &     \fullcirc    &         &         &         &         &                               &    \fullcirc                           &                               &                               &                               \\
filtering     &    & \cite{noauthor_natisand_nodate}         &       \cite{ebpf}      & \fullcirc       & \fullcirc       &    \halfcirc     &         &         &                              &                               &                               &             \fullcirc                  &                               \\
virtualization     &          &       \cite{sec-ci}            &    \cite{intel-vtx, amd-svm}   &         &    \fullcirc     &        \fullcirc  &       &         &                               &                               &                               &              \fullcirc                 &                               
\\[0.2em]  
\hline
\textbf{\rule{0pt}{1.2em}Steps}                  &                   &       &            & \textbf{S1} & \textbf{S2} & \textbf{S3} &     &     &                               &                               &                               &                               &                               \\
deterministic steps      &  \checkmark  &  \cite{verifiability, dettrace, guix} &             &  \fullcirc     &         &         &         &         &                               &                               &                                 &                               &         \fullcirc                      \\
step transparency   & \checkmark            &  \cite{TorresArias2019}   &       \cite{slsa, spdx, in-toto, trivy, sbom}          &        & \fullcirc       &  \halfcirc   &   &   \halfcirc      &     \fullcirc    &                               &                             &                               &                                          \\
step consensus  & \checkmark             & \cite{chainiac, Policy-Transparency}          &       \cite{scitt}    &        & \fullcirc       &   \fullcirc      &         &   \fullcirc      &       \fullcirc                          &                            &                               &                            &                  \\

static analysis  &    &   \cite{wu_your_nodate,meng_identifying_2017}            &            &   \fullcirc      &         &         &         &         &                               &                               &                               &                               &              \halfcirc                 \\
ML-based analysis  &   &  \cite{han_sigl_nodate}               &            &     \fullcirc    &         &         &         &         &                               &                               &                            &                               &           \fullcirc                    
\\[0.2em]
\hline
\textbf{\rule{0pt}{1.2em}Supply Chain Structure}    &               &     &              & \textbf{T1} & \textbf{T2} &     &     &     &                               &                               &                               &                               &                               \\
supply chain layout integrity & \checkmark  & \cite{TorresArias2019}           &    \cite{in-toto}       & \fullcirc       &         &         &         &         &          \fullcirc                      &                              &                               &                               &                             \\
supply chain reproducibility & \checkmark &   \cite{verifiability}               &         &         &    \fullcirc       &         &         &         &                               &                               &                               &                               &              \fullcirc                 
\\[0.2em] 
\hline
\end{tabular} \label{tab:sok-table2}
\end{table*}

We present a version of our proposed software supply chain defenses taxonomy that lists all 76 papers and industry sources we systematized in table~\ref{tab:sok-table2}.

%\input{related}
%%%%%%%%%%%%%%%%%%%%%%%%%%%%%%%%%%%%%%%%%%%%%%%%%%%%%%%%%%%%%%%%%%%%%%%%%%%%%%%%
%%%%%%%%%%%%%%%%%%%%%%%%%%%%%%%%%%%%%%%%%%%%%%%%%%%%%%%%%%%%%%%%%%%%%%%%%%%%%%%%

%%  LocalWords:  endnotes includegraphics fread ptr nobj noindent
%%  LocalWords:  pdflatex acks

\end{document}